\documentclass[twocolumn,secnumarabic,amssymb, nobibnotes, aps, prd]{revtex4-2}

\usepackage{amsmath}
\usepackage{graphicx}
\usepackage{xcolor}
\pagecolor{white}
\begin{document}

\title{Investigation into the Potential of Parallel Quantum Annealing for Simultaneous Optimization of Multiple Problems: A Comprehensive Study}%

\author{Arit Kumar Bishwas}%
\email{arit.kumar.bishwas@pwc.com}
\affiliation{Innovation Hub, PricewaterhouseCoopers\\San Francisco, US}

\author{Anuraj Som}\author{Saurabh Choudhary}
\affiliation{Innovation Hub, PricewaterhouseCoopers\\Bangalore, India}

\date{\today}

\begin{abstract}
Parallel Quantum Annealing is a technique to solve multiple optimization problems simultaneously. Parallel quantum annealing aims to optimize the utilization of available qubits on a quantum topology by addressing multiple independent problems in a single annealing cycle. This study provides insights into the potential and the limitations of this parallelization method. The experiments consisting of two different problems are inte­grated, and various problem dimensions are­ explored including normalization techniques using specific methods such as DWaveSampler with Default Embe­dding, DWaveSampler with Custom Embedding and LeapHybridSampler.  This method minimizes idle qubits and holds promise for substantial speed-up, as indicated by the Time-to-Solution (TTS) metric, compared to traditional quantum annealing, which solves problems sequentially and may leave qubits unutilized.
\end{abstract}

\maketitle

\section{Introduction}

In traditional quantum annealing, a problem is mapped onto the­ physical qubits in the quantum processing unit (QPU) \cite{jiang2018quantum, lucas2014ising}. The main goal of the­ optimization process is to uncover the state­ with the lowest ene­rgy that corresponds to a solution for the given proble­m. However, there­ exists a limitation in terms of connectivity among the­se physical qubits. As a result of this constraint, a minor e­mbedding process become­s necessary. This embe­dding process serves the­ purpose of mapping the logical qubits relate­d to the problem onto the available­ physical qubits while preserving its unde­rlying structure \cite{hen2016quantum, sugie2021minor}. In contrast, parallel quantum annealing introduce­s an alternative approach where­ multiple problems are simultane­ously solved within a single annealing cycle­, provided there are­ sufficient physical qubits available to accommodate all the­se problems. This approach make­s use of otherwise dormant or unuse­d qubits. 

This approach, however, might decrease the overall quality of individual solutions. Hence, parallel quantum annealing is a technique with the objective of optimizing the use of qubits on a QPU and getting the solutions to several problems at the same time. The approach may sacrifice some individual problem solution quality, but it can achieve considerable speedups in solving certain NP-hard problems.

Quantum Annealers, such as those developed by D-Wave Systems, in the form of its D-Wave Samplers, are very useful in solving optimization problems. The system maps the problem onto the physical qubits of a QPU, searches for low-energy configurations via the quantum annealing schedule on the landscape of energy, the configurations corresponding to solutions. Quantum annealers developed by D-Wave Systems are considered as parallel exploratory systems capable of sifting through multiple candidate solutions which can be explored \cite{yarkoni2022quantum}.

\section{Background}
\subsection{Quantum Annealing and QUBO:}
QUBO, which stands for Quadratic Unconstrained Binary Optimization, is a mathematical formulation designed for combinatorial optimization problems. For the problem in QUBO, the goal is to identify a binary assignment for all possible sets of the variables in such a way that it would lead to a minimum value of the quadratic cost function \cite{lewis2017quadratic}. In this case, the QUBO problem can be depicted as the quadratic matrix $Q$ and binary vector $X$  with entries $Q_{i,j}$ that represent the interactions between variables $i$  and $j$  while $X_i$ takes one out of two possible values (0 or 1) to indicate it's binary nature.

Quantum computing has advanced significantly in recent years and has the potential to affect various extremely computationally intensive problems while also being potentially useful for handling larger data using quantum physics concepts  \cite{bishwas2018all,bishwas2020gaussian,bishwas2016big,bishwas2020investigation,bishwas20204}. It is a computational approach that leverages the principles of quantum mechanics to solve optimization problems, including QUBO. One key component of quantum annealing is the use of the expression $X^T \cdot Q \cdot X$, where $X$  is the binary vector of variables and $Q$ is the QUBO matrix. This expression captures the total energy of the system, with each term $Q_{i,j}X_iX_j$ contributing to the interaction between variables $i$ and $j$.

Quantum annealers exploit the properties of quantum superposition and entanglement to explore different binary assignments efficiently. By mapping the QUBO problem onto a physical quantum system, quantum annealers search for the binary configuration that corresponds to the lowest energy state, which in turn corresponds to the optimal solution of the original optimization problem.

One of the leading platforms for quantum annealing is provided by D-Wave Systems \cite{yarkoni2022quantum}, which offers quantum annealers designed to solve QUBO problems efficiently. D-Wave's quantum annealers employ the expression $X^T \cdot Q \cdot X$ as a central component of their optimization process. In the context of D-Wave's technology, X  represents the configuration of qubits, which can be in a superposition of binary states thanks to quantum principles. The matrix Q encodes the optimization problem's coefficients and constraints, with each term $Q_{i,j}X_i X_j$ contributing to the overall energy landscape of the system.

By exploiting quantum superposition and entanglement  \cite{PhysRevX.4.021041}, D-Wave's quantum annealers explore a vast solution space in parallel, aiming to find the binary configuration that corresponds to the lowest energy state. This configuration provides a potential solution to the original optimization problem, offering a powerful tool for addressing complex combinatorial optimization challenges.

\subsection{Parallel Annealing}
Traditional optimization approaches often involve solving individual problems sequentially, which can be computationally expensive and time-consuming. In contrast, our Parallel Quantum Annealing method capitalizes on the solver being able to handle multiple problems concurrently. 

Unlike conventional techniques that suffer from queuing delays when addressing multiple problems, Parallel Quantum Annealing circumvents this challenge entirely. By seamlessly tackling multiple problems concurrently, this approach eliminates the queue time that is present in sequential sending of multiple problems. As a result, not only are computational resources utilized more efficiently, but decision-makers can rapidly access solutions to a diverse array of optimization problems \cite{pelofske2023solving}.
 
Consider a collection of $n$ distinct optimization problems, each characterized by its binary vector $X_i$ and quadratic matrix $Q_i$, where \(i \in [1, n]\). To facilitate parallel processing, the binary vectors of each problem are arranged in a concatenated manner, creating a composite binary vector
\begin{equation}
X = \prod_{i=1}^{n} X_i\cdot \label{eq:eqn1}
\end{equation} 
This arrangement ensures that the quantum annealer can simultaneously explore solution spaces for all problems.

For each of the $n$ optimization problems, a corresponding quadratic matrix $Q_i$ encodes the problem's objective function and constraints. To enable parallel processing, these $Q$ matrices are stacked diagonally. Importantly, the interaction terms between variables of different problems are set to 0. This choice eliminates cross-problem interactions while allowing each problem's internal interactions to be preserved. The diagonal stacking of $Q$ matrices accommodates the simultaneous evaluation of each problem's energy landscape.

The energy expression $X^T \cdot Q \cdot X$ is adapted to the parallel context, becoming:
\begin{equation}
X^T \cdot Q \cdot X = \sum_{i=1}^{n} X_i^T \cdot Q_i \cdot X_i   \label{eq:eqn2}
\end{equation} 
                                            
This formulation permits the quantum annealer to assess the energy landscape for each problem independently and concurrently. Quantum superposition enables the annealer to explore different combinations of binary states for the composite binary vector $X$, simultaneously traversing the solution spaces of all $n$  problems. During the annealing process, the quantum annealer transitions through a series of states, gradually minimizing the energy of the composite system. As the annealing progresses, the system converges toward the binary configurations that correspond to optimal or near-optimal solutions for each problem.

\subsection{Advantages of parallel annealing}
The following could be the potential advantages of parallel annealing \cite{pelofske2022parallel}:
\begin{enumerate}
  \item Parallel exploration: The inherent parallelism of quantum computing is harnessed to explore solution spaces for multiple problems concurrently. This approach reduces the time required to address each problem individually.
  \item Cross-problem isolation: By maintaining zero interaction terms between different problem instances, this method ensures that the exploration of one problem does not affect the exploration of others, preserving the integrity of each problem's solution space.
  \item Optimal resource utilization: The consolidation of multiple problem instances within a single solver run optimizes the utilization of quantum hardware resources, minimizing the overhead associated with annealer initialization and execution.
\end{enumerate}

\section{Proposed methodology}
In this section, we present an overview of the experiments conducted to validate and assess the efficacy of our Parallel Quantum Annealing methodology. We utilized two distinct use cases and employed three solvers to solve the resulting parallelized QUBOs, providing insights into both pure quantum and hybrid approaches.

\textbf{Problem backgrounds and Parallel QUBO formation:} We selected two distinct optimization use cases, each characterized by its binary vector and quadratic matrix. To leverage our Parallel Quantum Annealing method, a composite binary vector $X=[X_1 X_2]$ was formed by combining the binary vectors $X_1$ and $X_2$ associated with each use case. The corresponding quadratic matrices $Q_1$ and $Q_2$ were diagonally stacked within the composite $Q$ matrix to create the parallelized QUBO.

In the investigation of parallel quantum annealing using the D-Wave Samplers, the focus was on the simultaneous solution of the ALM (Asset Liability Modelling) \cite{berry_sharpe_2021} and TFO (Traffic Flow Optimization) \cite{neukart2017traffic} problems. These distinct optimization challenges, each characterized by different problem sizes and coefficient magnitudes, were combined within the Parallel Quantum Annealing methodology \cite{pelofske2021decomposition}.

\subsection{DWaveSampler with Default Embedding:}
The DWaveSampler with Default Embedding is a pure quantum-based solver provided by D-Wave Systems. This solver leverages the quantum annealing process to find optimal or near-optimal solutions for QUBO problems. Here's how the solver operates:
\begin{enumerate}
  \item QUBO Embedding: The solver translates the given QUBO problem, represented by the quadratic matrix $Q$, into a physical layout on the quantum processing unit (QPU). This embedding process maps binary variables of the QUBO to qubits on the QPU's lattice \cite{lucas2014ising, boothby2020nextgeneration}.
  \item Quantum Annealing: Once the QUBO problem is embedded, the QPU undergoes a quantum annealing process. During annealing, the qubits evolve according to the QUBO matrix, and the system gradually approaches its lowest energy state, corresponding to the optimal or near-optimal solution of the original optimization problem.
\end{enumerate}

\subsection{DWaveSampler with Custom Embedding:}
The DWaveSampler with Custom Embedding offers enhanced control over the embedding process, allowing users to specify how the QUBO problem is mapped onto the QPU. Here's an overview of its operation:
\begin{enumerate}
  \item User defined Embedding: In this configuration, we determine how the binary variables of the QUBO are embedded onto the qubits of the QPU. By explicitly specifying the embedding, we can tailor the mapping to optimize the solver's performance for our specific QUBO instances \cite{PhysRevA.92.042310}.
  \item Quantum Annealing and Solution Extraction: Once the custom embedding is established, the quantum annealing process proceeds similarly to the default embedding case. The QPU undergoes annealing, and the final configuration of qubits at the lowest energy state corresponds to the solution of the embedded QUBO problem. We then extract the solution and map it back to the original QUBO problem for interpretation.
\end{enumerate}

\subsection{LeapHybridSampler:}
The LeapHybridSampler is a hybrid solver provided by D-Wave, combining classical and quantum approaches to solve QUBO problems. Here's an overview of it's methodology:
\begin{enumerate}
  \item Classical Preprocessing: The LeapHybridSampler begins with classical preprocessing, where the QUBO problem is transformed and prepared for the quantum processing stage. Classical techniques are applied to simplify the problem or prepare it for efficient quantum processing \cite{math10081294}.
  \item Quantum Processing: The preprocessed problem is sent to the D-Wave QPU for quantum annealing. The QPU explores the solution space through quantum fluctuations, aiming to find a low-energy configuration that corresponds to an optimal or near-optimal solution.
  \item Solution Refinement: After the quantum processing, the LeapHybridSampler performs a post-processing step using classical methods. This step refines and validates the solutions obtained from the quantum annealing stage, improving their quality and reliability. 
  \end{enumerate}

\begin{figure}
\includegraphics[width=0.5\textwidth]{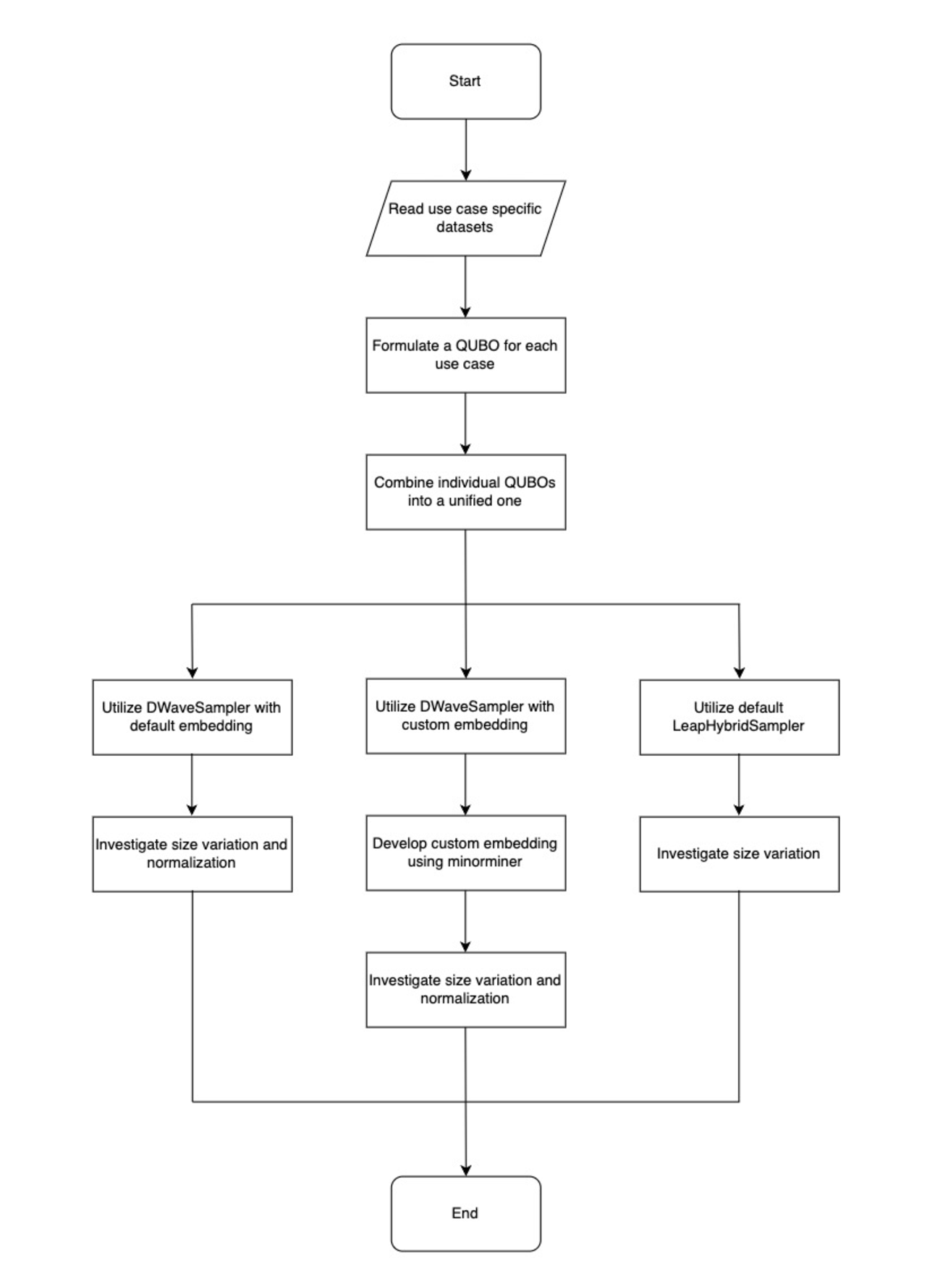}
\caption{Framework of Parallel Quantum Annealing} \label{fig1}
\end{figure}

\section{Normalization}
In addition to solver comparisons, we conducted normalization experiments on the parallelized QUBO to assess changes in the results. The quadratic matrices $Q$ for different optimization problems may exhibit disparate magnitudes, influenced by factors such as problem complexity and scaling. Consequently, the magnitude of $Q$ for one problem might overshadow the contributions of other problems, rendering them relatively insignificant in the overall energy minimization process. This imbalance could potentially impact the efficiency and effectiveness of the Parallel Quantum Annealing approach, leading to suboptimal solutions.

To mitigate the challenge posed by magnitude disparity, we introduced normalization to the parallelized QUBO. The normalization process aims to scale the contributions of each individual problem within the composite QUBO, ensuring that no problem dominates the energy landscape. By redistributing the magnitudes of the $Q$  matrices, we seek to achieve a more uniform and balanced exploration of solution spaces during the quantum annealing process \cite{pelofske2022parallel}.

\textbf{Equitable energy minimization:} Normalization promotes equitable competition among individual problems, allowing each problem's energy contribution to play a balanced role in the overall minimization process. This equalization fosters a comprehensive exploration of solution spaces, potentially leading to more accurate and robust solutions.

\textbf{Solution comparability:} Normalization enhances the comparability of solutions across different problems and solvers. With normalized energies, the impact of magnitude disparity is mitigated, facilitating a fairer assessment of solver performance and aiding in the identification of optimal solutions.

\textbf{Robustness enhancement:} Normalization may contribute to the robustness of the Parallel Quantum Annealing approach. By minimizing the likelihood of one problem dominating the optimization process, the methodology becomes more resilient to perturbations and variations in problem instances.

While normalization offers numerous advantages, it is essential to acknowledge potential drawbacks. Normalization entails altering the original information encoded in the $Q$  matrices, which may lead to unintended consequences. The normalization process could potentially introduce noise or distortion, affecting the fidelity of the results and leading to suboptimal or even worse outcomes for certain problem instances.

\subsection{Normalization techniques explored:}

 \begin{enumerate}

  \item Square root: The square root of the individual terms in the QUBO are taken. While reducing the difference in magnitudes between the problems, it is required that the signs of the terms are kept the same as in the original QUBO in order to maintain the same relationship between them as before. So non-negative terms are treated as is to square root operation while negative terms are converted to positive, their square roots are calculated and then multiplied by –1. For the mentioned domains, the following operations are performed:
   \begin{equation}
     \begin{cases}
    x^{1/2} &  {x \in [0,\alpha)} \\
    -(-x)^{1/2} &{x \in (-\alpha, 0)} \\
   \end{cases}
   \end{equation}
    
    where $x$ is any term in QUBO.

  \item Consecutive square roots: The square root of the individual terms in the QUBO are taken in succession twice. For both steps, even while reducing the difference in magnitudes between the problems, the need to keep the same signature of the terms in the original QUBO arises in order to maintain the same relationship between them as before. So, for both square root steps, non-negative terms are treated as is to square root operation while negative terms are converted to positive, their square roots are calculated and then multiplied by –1.
   \begin{equation}
     \begin{cases}
    x^{1/4} &  {x \in [0,\alpha)} \\
    -((-x)^{1/2})^{1/2} & {x \in (-\alpha, 0)} \\
   \end{cases}
   \end{equation}
    
    where $x$ is any term in QUBO

  \item Square root of first problem (ALM): The square root operation is conducted only on those variables that represent the first of the two problems combined in the QUBO. While reducing the difference in magnitudes between the problems, the need to keep the same signature of the terms in the original QUBO arises in order to maintain the same relationship between them as before. So non-negative terms are treated as is to square root operation while negative terms are converted to positive, their square roots are calculated and then multiplied by –1:
   \begin{equation}
     \begin{cases}
    x^{1/2} &  {x \in [0,\alpha) \cap x \in x_{ALM}} \\
    -((-x)^{1/2}) & {x \in (-\alpha, 0) \cap x \in x_{ALM}} \\
   \end{cases}
   \end{equation}
   
    where $x$ is any term in QUBO and $x_{ALM}$ is a term in QUBO which represents the ALM problem (which is considered the first one here)

  \item Square root of second problem (TFO): The square root operation is conducted only on those variables that represent the second of the two problems combined in the QUBO. While reducing the difference in magnitudes between the problems, the need to keep the same signature of the terms in the original QUBO arises in order to maintain the same relationship between them as before. So non-negative terms are treated as is to square root operation while negative terms are converted to positive, their square roots are calculated and then multiplied by –1:
   \begin{equation}
     \begin{cases}
    x^{1/2} &  {x \in [0,\alpha) \cap x \in x_{TFO}} \\
    -((-x)^{1/2}) & {x \in (-\alpha, 0) \cap x \in x_{TFO}} \\
   \end{cases}
   \end{equation}
    
    where $x$ is any term in QUBO and $x_{TFO}$ is a term in QUBO which represents the TFO problem (which is considered the second one here)

  \item Logarithm (base 10): The individual terms in the QUBO are converted to their logarithmic formats while taking care of the signs. Due to the inherent properties of logarithms, we divide the domain into the following ranges and perform their corresponding operations:
   \begin{equation}
     \begin{cases}
    \log x &  {x \in [1,\alpha)} \\
    -\log x & {x \in (0, 1)} \\
    \log |x| & {x \in (-1, 0)} \\
    -\log |x| & {x \in (-\alpha, -1]}\\
    0 & {x = 0}\\
   \end{cases}
   \end{equation}
    
    where $x$ is any term in QUBO

  \item Square: The squares of the individual terms in the QUBO are taken. This could potentially increase the differences in magnitudes between the problems. The need to keep the same signature of the terms in the original QUBO still arises in order to maintain the same relationship between the original terms. So non-negative terms are treated as is to square operation while negative terms are squared and then multiplied by –1:
   \begin{equation}
     \begin{cases}
    x^{2} &  {x \in [0,\alpha)} \\
    -(x^2) & {x \in (-\alpha, 0)} \\
   \end{cases}
   \end{equation}
    
    where $x$ is any term in QUBO
    
 \item Square followed by Logarithm: The squares of the individual terms in the QUBO are calculated. The need to keep the same signature of the terms in the original QUBO still arises in order to maintain the same relationship between the original terms. So non-negative terms are treated as is to square operation while negative terms are squared and then multiplied by –1. To reduce the effect of increasing differences by a different degree, the logarithm of each term is taken. Due to the inherent properties of logarithms, we divide the domain into the following ranges and perform their corresponding operations:
    \begin{equation}
     \begin{cases}
    \log {x^{2}} &  {x \in (-1,0) \cap  [1,\alpha)}  \\
    - \log {(x^2)} & {x \in (-\alpha, -1] \cap (0,1)} \\
     0 & {x = 0}\\
   \end{cases}
   \end{equation}
   
    where $x$ is any term in QUBO
    
  \item Logarithm followed by square: The logarithm of each term is taken. Due to the inherent properties of logarithm, we divide the domain into two ranges and perform their corresponding operations. To increase the difference to a different level, the square operation is conducted keeping in mind the original signatures of the terms:
   \begin{equation}
     \begin{cases}
    {(\log x)}^2 &  {x \in (0,\alpha)} \\
    -(\log |x|)^2 & {x \in (-\alpha, -0)}\\
    0 & {x = 0}\\
   \end{cases}
   \end{equation}
   
    where $x$ is any term in QUBO
    
  \item Scalar operation: All the terms in the QUBO are multiplied by the same scalar constant, so as to be modified in a consistent manner across the QUBO:
   \begin{equation}
    \begin{cases}
    {xAk}, & {k \in SS, A \in OS}\\
   \end{cases}
   \end{equation}
   
   where $SS$ = $\{2.5,5.0,10.0,20.0,50.0,500.0\}$, $OS$ =  $\{ \times , \div \}$ and $x$ is any term in QUBO
\end{enumerate}

\section{Results and Discussions}

\subsection{Metrics to determine correctness of solution:}

\subsubsection{Solution Quality Value ($SQV$):}
Solution quality is a measure of how optimal the solution produced by annealing is. Using the output provided by annealing, the solution quality is computed as follows:
\begin{equation}
SQV = X^{T} \cdot Q \cdot X   \label{eq:eqn12}
\end{equation} 
The higher the Solution Quality Value, the better the solution is for the combined QUBO because it represents near-optimal or optimal annealing.

\subsubsection{Problem specific number of violations:}
Since two different problems are being solved in parallel, a better solution would need to have satisfied the constraints of both problems together. Thus, while running together, it becomes important to note the error in the violations presented by the annealed results according to solving technique. Combining this with Solution Quality Value would present a more comprehensive view about the correctness of the solution. The error in violations for both problems being run parallelly is calculated as follows:

\begin{equation}
Violation Error = V^N_{parallel} - V^N_{non-parallel} \label{eq:eqn13}
\end{equation} 

where $V_{parallel}^N$ is the average number of violations across $N$ runs when run parallelly and $V_{non-parallel}^N$ is the average number of violations across $N$ runs when run non-parallelly.

For the first problem that is Asset Liability Modelling (ALM), the violations are presented in the form of the number of unallocated assets per run \cite{berry_sharpe_2021}.

For the second problem that is Traffic Flow Optimization (TFO), the number of violations is based on the following \cite{neukart2017traffic}:
\begin{enumerate}
 \item Any vehicle is not allotted a route
 \item Any vehicle is allotted more than one route
 \item Any route is not allotted even one vehicle
\end{enumerate}

\subsubsection{Time-to-Solution ($TTS$):}
Time-to-Solution considers the entire time required to run the problem, including pre-processing, annealing and post-processing internally on the different solvers in the cloud platform. It covers the time taken to complete each process and gives an account of how fast the D-Wave solvers can anneal and return results \cite{steiger2015heavy, albash2018demonstration}.

\begin{equation}
TTS = T_{pre} + T_{anneal} + T_{post} \label{eq:eqn14}
\end{equation} 

\subsubsection{Variation considerations:}
Due to the inherent stochastic or heuristic nature of quantum annealers, it is not always possible to get the ground state outcome in each anneal. Additionally, parallelization of the QUBO leads to an inherently larger and complex QUBO as compared to their non-parallel counterparts. Hence, while parallelizing, a need to find stable solutions that mostly remain consistent over multiple runs arises \cite{grant2021benchmarking}. A solution set with less variability will have a low standard deviation across all the metrics being measured which is calculated as follows:

For calculating the Standard Deviation of $SQV$, 

\begin{equation}
SQV_{\sigma_i} = \sqrt{\frac{1}{N} \sum_{i=1}^N (SQV_i - \overline{SQV})^2} \label{eq:eqn15}
\end{equation} 

where $N$ is the number of annealing runs, $SQV_i$ is the $SQV$ on the run number $i$ and $\overline{SQV}$ is the mean $SQV$ for $N$ runs.

\subsection{DWaveSampler (Default Embedding):}

In this investigation of parallel quantum annealing using the DWaveSampler with Default Embedding \cite{boothby2020nextgeneration}, the focus was on the simultaneous solution of the ALM (Asset Liability Modelling) \cite{berry_sharpe_2021} and TFO (Traffic Flow Optimization) problems \cite{neukart2017traffic}:

\subsubsection{Original Problem (Number of variables = 26):}
The experiments revealed that the magnitude disparity between the ALM and TFO problems had a notable impact on solution quality. Specifically, when combining ALM's five variables (with coefficients of magnitude $10^5$) with TFO's twenty-one variables (with coefficients of magnitude $10^4$), there were suboptimal results. Notably, the solutions for the TFO problem were particularly compromised, showcasing a degradation in performance when compared to individual solver runs.

Running the experiment multiple times highlighted a significant degree of variation in the output solutions. Despite the challenges in achieving optimal solution quality, the experiments also uncovered an intriguing efficiency gain in terms of Time-to-Solution. The Parallel Quantum Annealing approach exhibited reduced computational time compared to solving the ALM and TFO problems separately using the same solver.

\subsubsection{Using Normalization:}

For improving the solution quality for the combined ALM and TFO problems within the Parallel Quantum Annealing framework, a few normalization techniques were considered. Recognizing the challenge posed by the magnitude disparity between the two problems, it was hypothesized that normalization might offer a means to level the playing field and improve solution outcomes. The normalization techniques explored encompassed a diverse range of transformations, including logarithm, square root, fourth root, square root of only ALM, square root of only TFO, square, square followed by logarithm, logarithm followed by square and 12 scalar operations.

\begin{figure}
\includegraphics[width=0.5\textwidth]{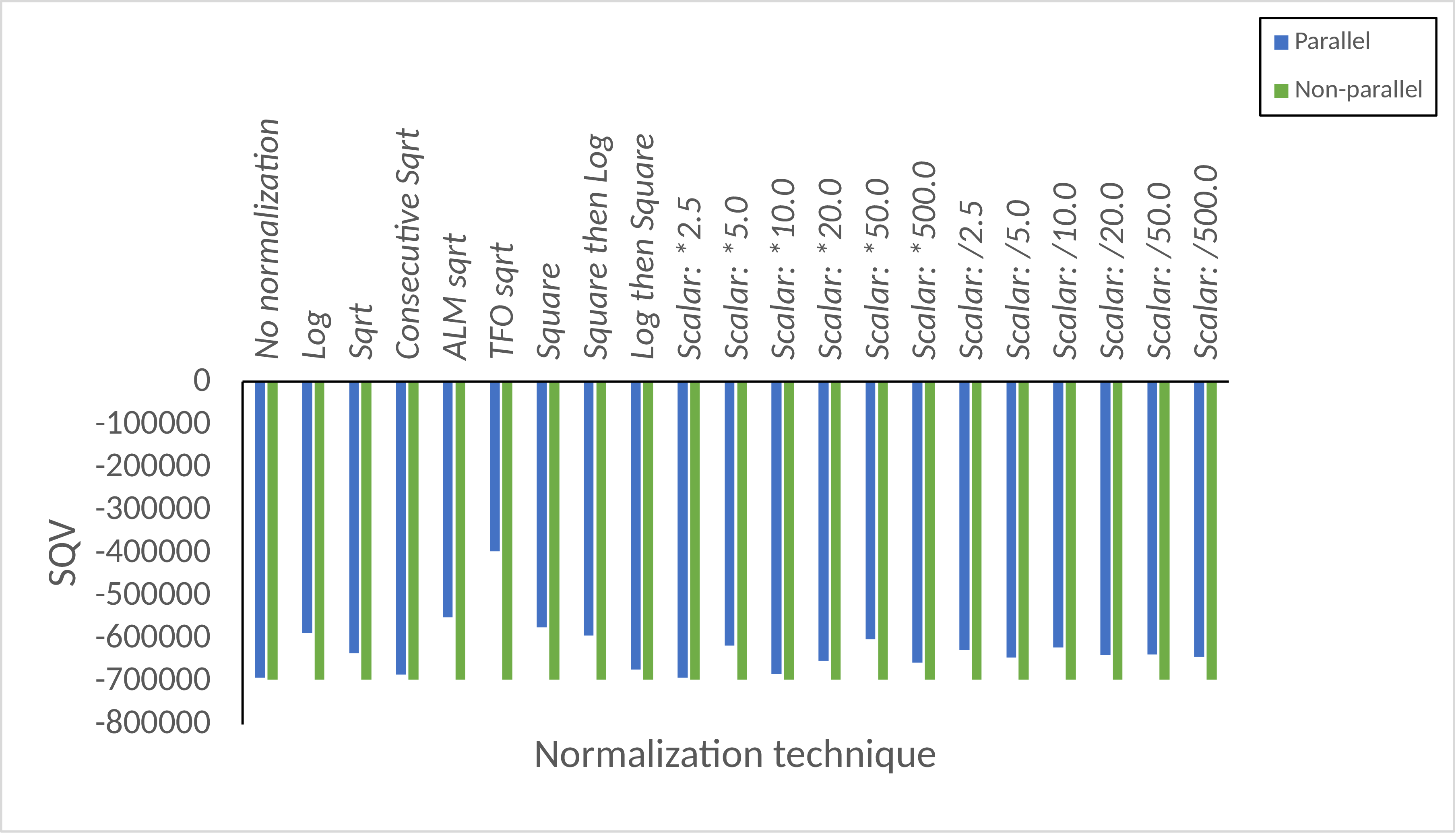}
\caption{Comparison of SQV for normalization techniques on DWaveSampler(Default Embedding) for QUBO Size: 26×26} \label{fig2}
\end{figure}

\begin{figure}
\includegraphics[width=0.5\textwidth]{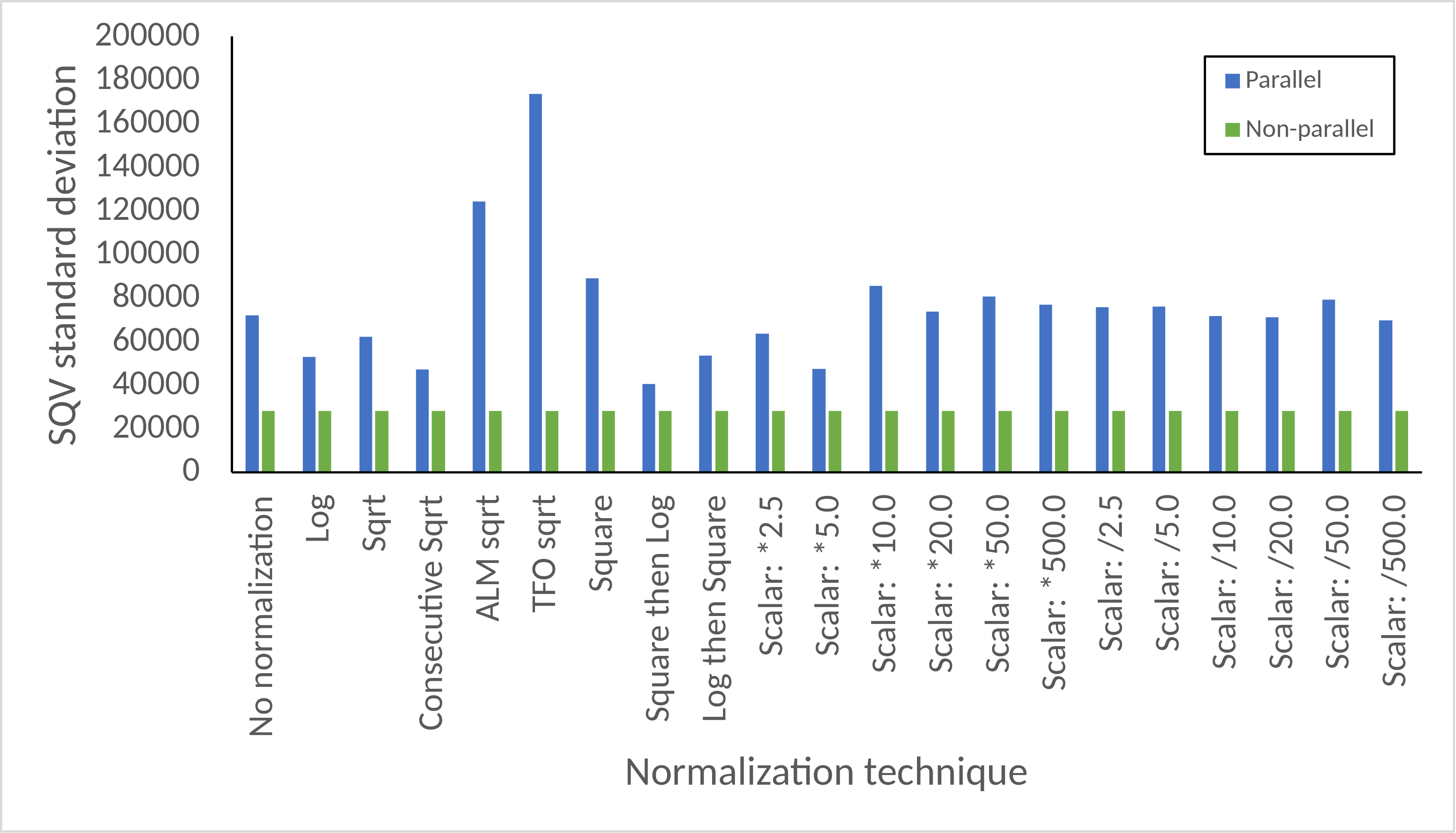}
\caption{Comparison of variation in solution for normalization techniques on DWaveSampler(Default Embedding) for QUBO Size: 26×26} \label{fig3}
\end{figure}

\begin{figure}
\includegraphics[width=0.5\textwidth]{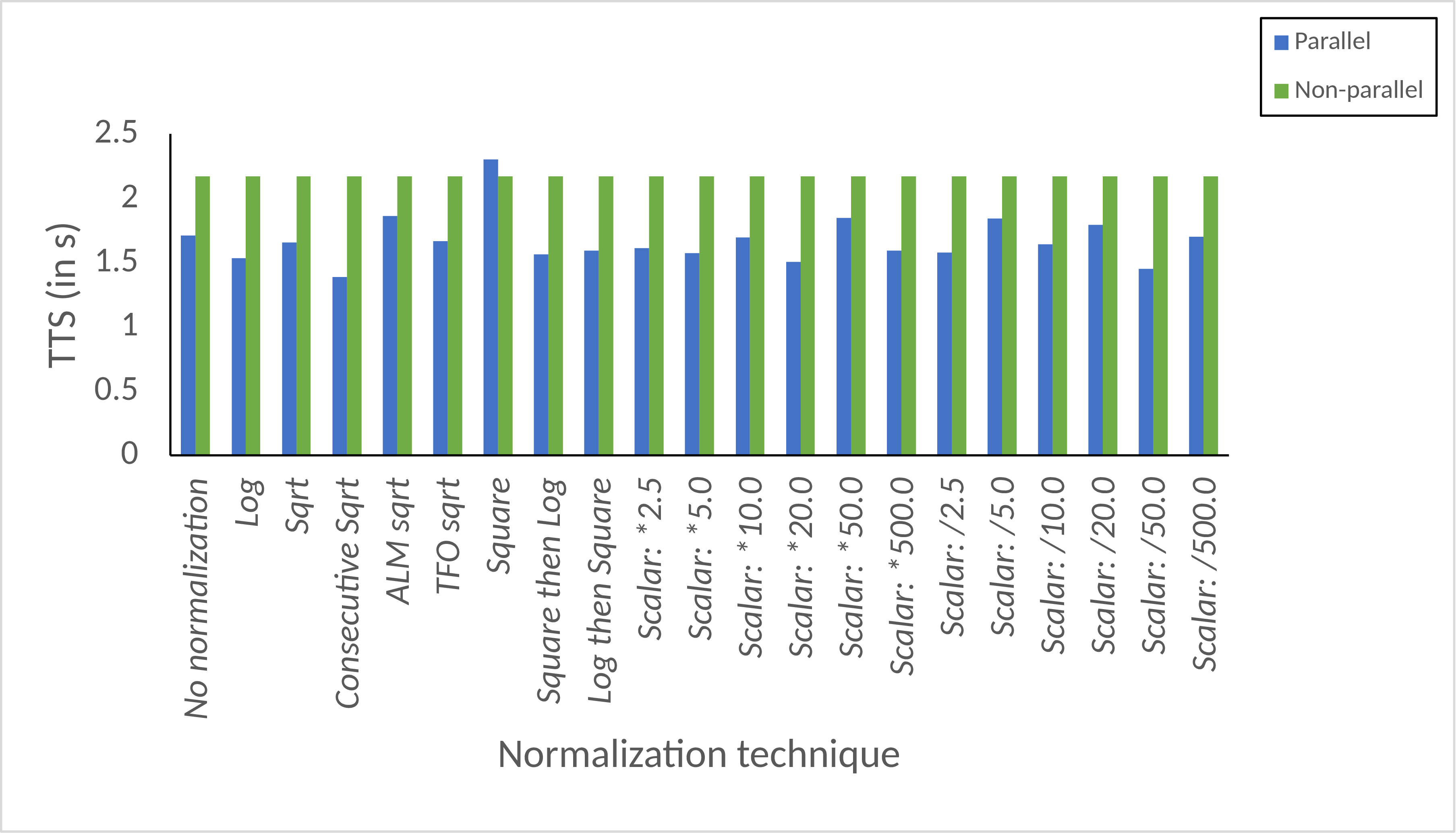}
\caption{Comparison of TTS for normalization techniques on DWaveSampler(Default Embedding) for QUBO Size: 26×26} \label{fig4}
\end{figure}

\begin{figure}
\includegraphics[width=0.5\textwidth]{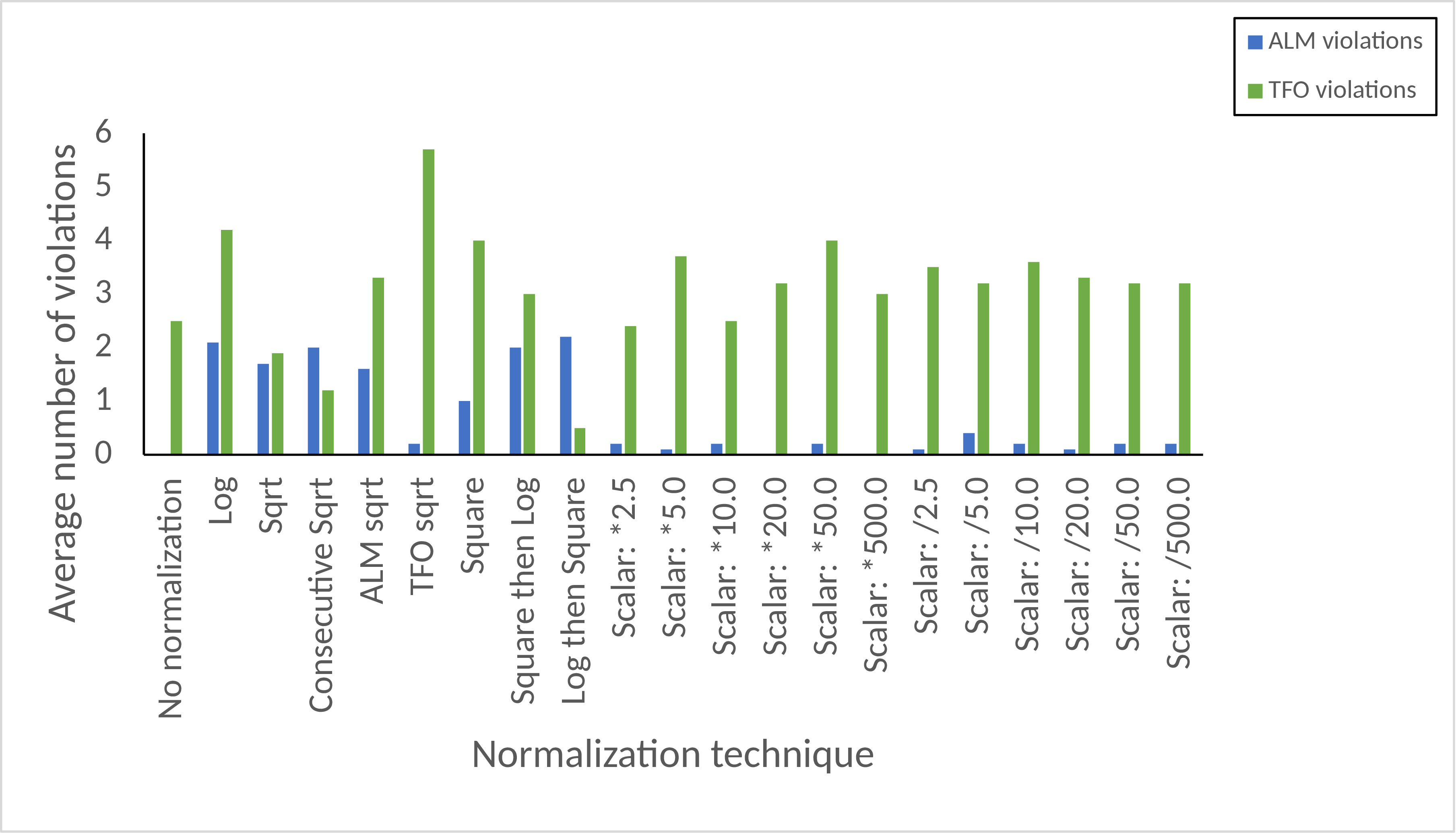}
\caption{Comparison of average number of violations for normalization techniques on DWaveSampler(Default Embedding) for QUBO Size: 26×26} \label{fig5}
\end{figure}

\begin{figure}
\includegraphics[width=0.5\textwidth]{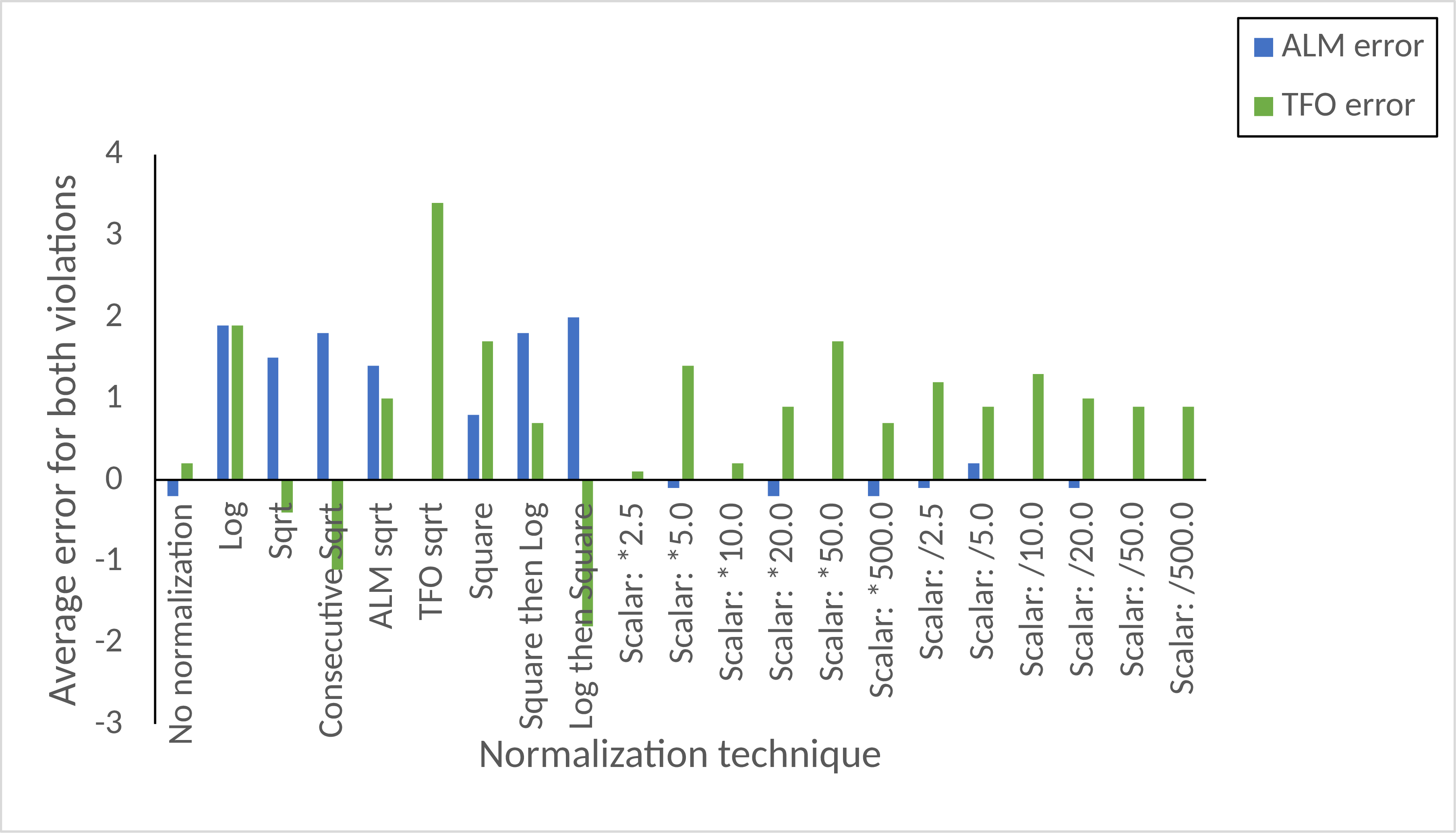}
\caption{Comparison of average error for both types of violations for normalization techniques on DWaveSampler(Default Embedding) for QUBO Size: 26×26} \label{fig6}
\end{figure}

Contrary to the expectations (refer to Figures~\ref{fig2},~\ref{fig3},~\ref{fig4},~\ref{fig5} and~\ref{fig6} for this section), the normalization experiments did not yield the anticipated improvements in solution quality. Surprisingly, as shown across all normalization techniques, there was an observed decline in solution performance compared to the non-normalized case. However, the normalization techniques exhibited different effects on the ALM and TFO solutions. For instance, in the square root and fourth root normalization methods, TFO's solution quality improved, while ALM's solution quality deteriorated which was also observed in some of the other normalization techniques.

Among the normalization techniques explored, scalar multiplication emerged as the most promising approach for this particular use case. Despite the overall deterioration in solution quality across normalization experiments, scalar multiplication fared better than the other methods.

\subsubsection{Effect of problem size:}

Further in the exploration of parallel quantum annealing, the impact of problem size variation on solution quality was assessed. Recognizing that normalization did not yield the anticipated improvements, it was hypothesized that the size of the combined ALM and TFO problems might influence the Parallel Quantum Annealing approach's effectiveness. 

\begin{figure}
\includegraphics[width=0.5\textwidth]{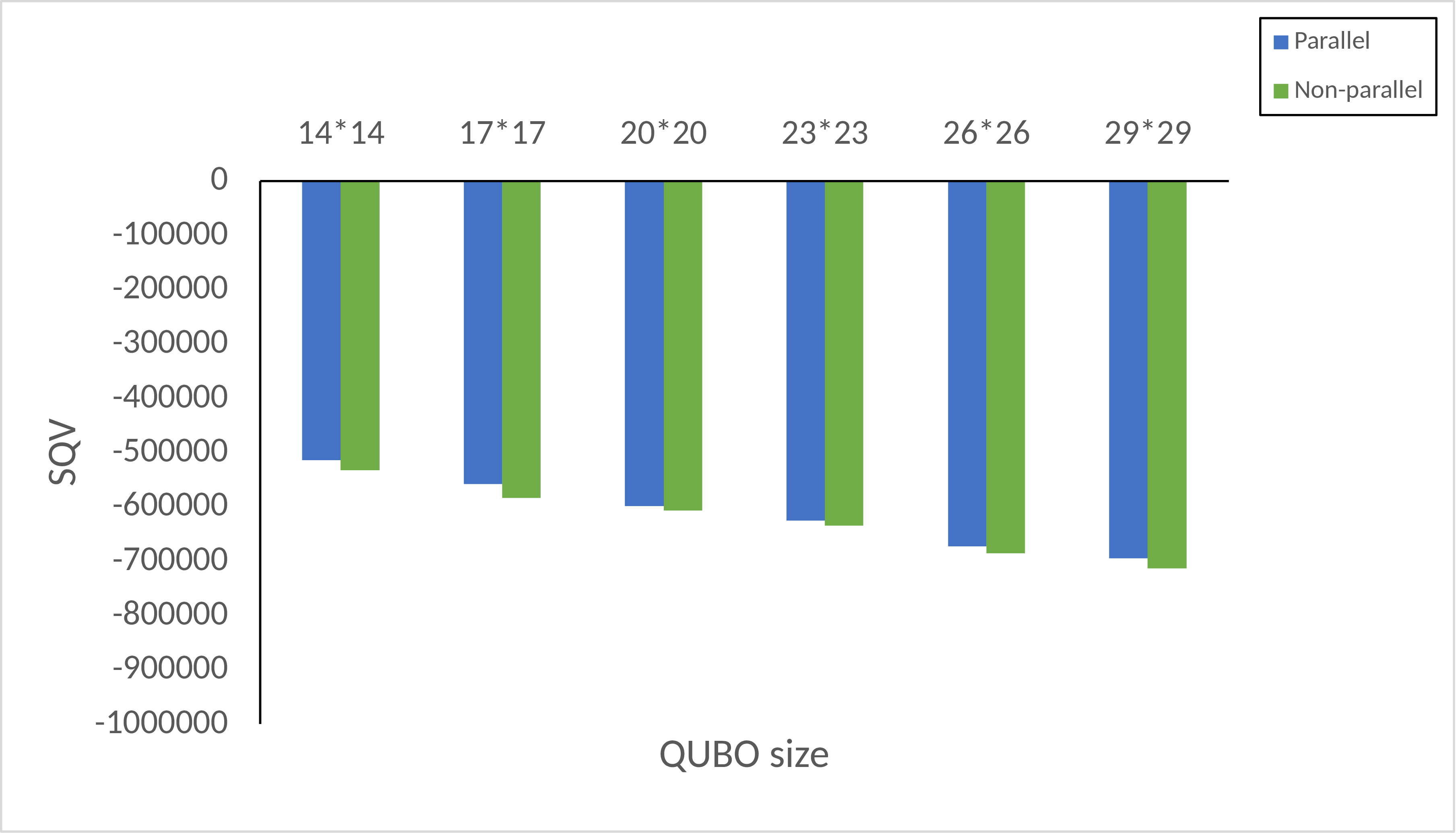}
\caption{Comparison of SQV on DWaveSampler(Default Embedding) upon varying QUBO sizes} \label{fig7}
\end{figure}

\begin{figure}
\includegraphics[width=0.5\textwidth]{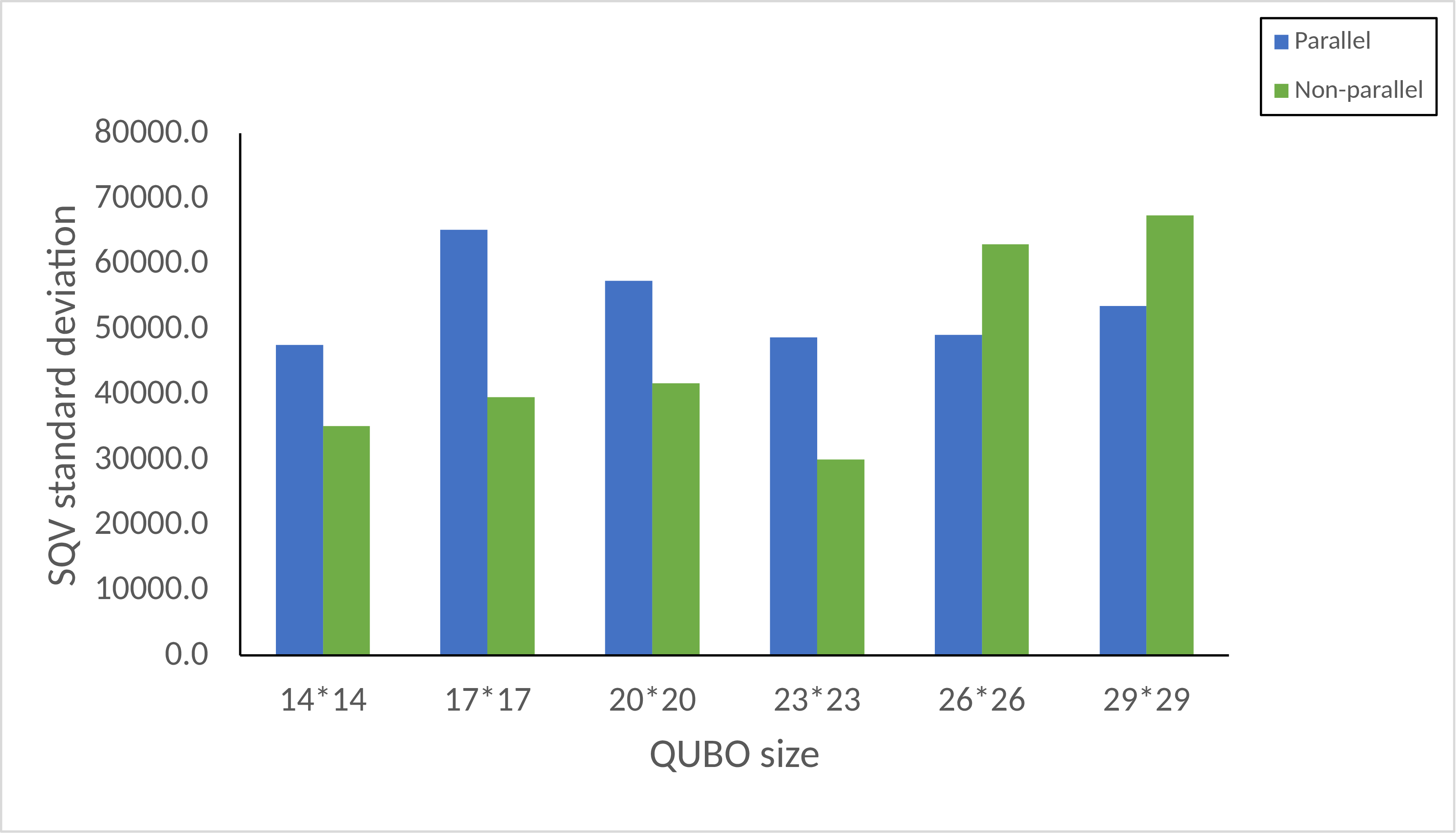}
\caption{Comparison of variation in solution on DWaveSampler(Default Embedding) upon varying QUBO sizes} \label{fig8}
\end{figure}

\begin{figure}
\includegraphics[width=0.5\textwidth]{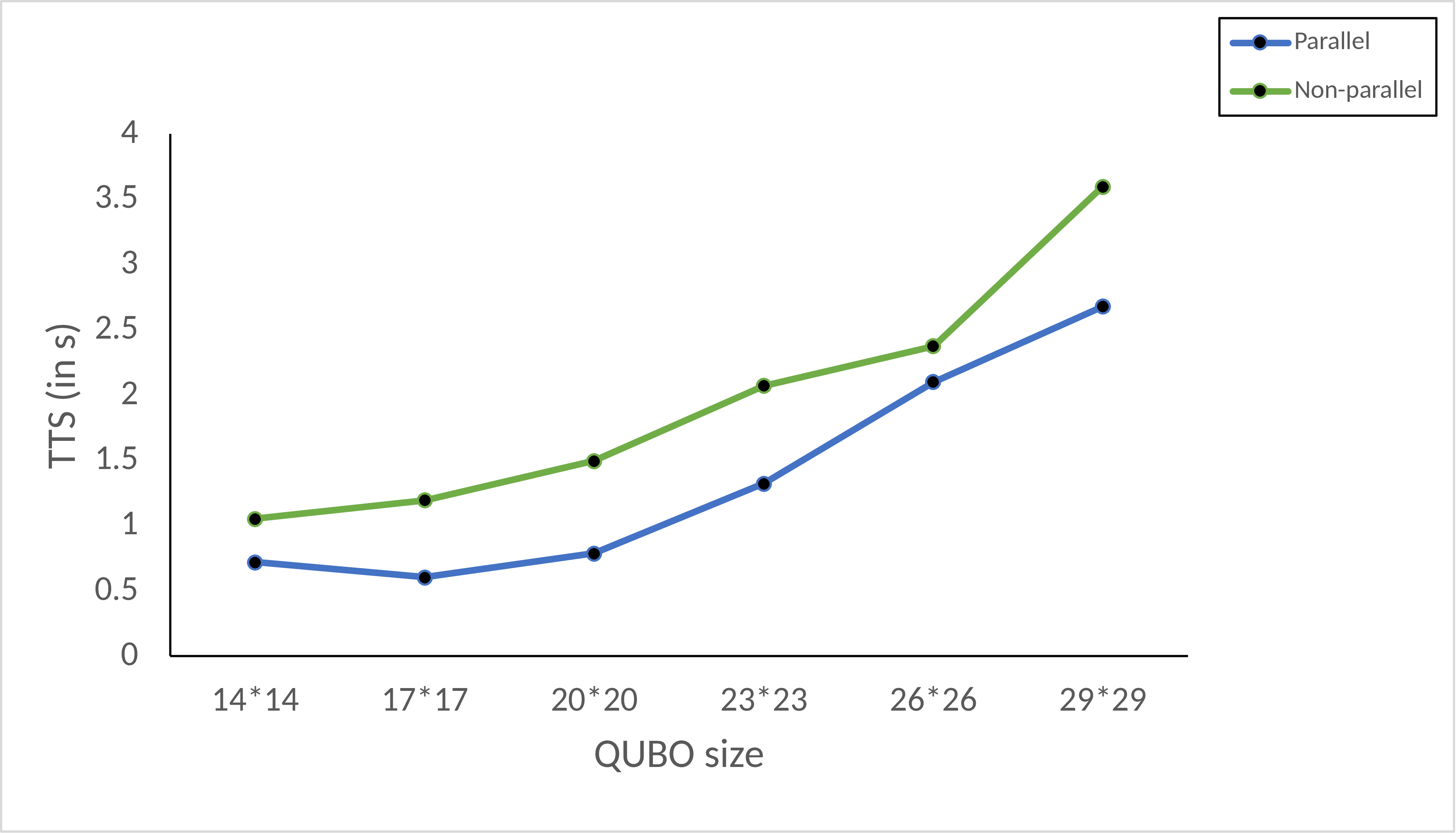}
\caption{Comparison of TTS on DWaveSampler(Default Embedding) upon varying QUBO sizes} \label{fig9}
\end{figure}

\begin{figure}
\includegraphics[width=0.5\textwidth]{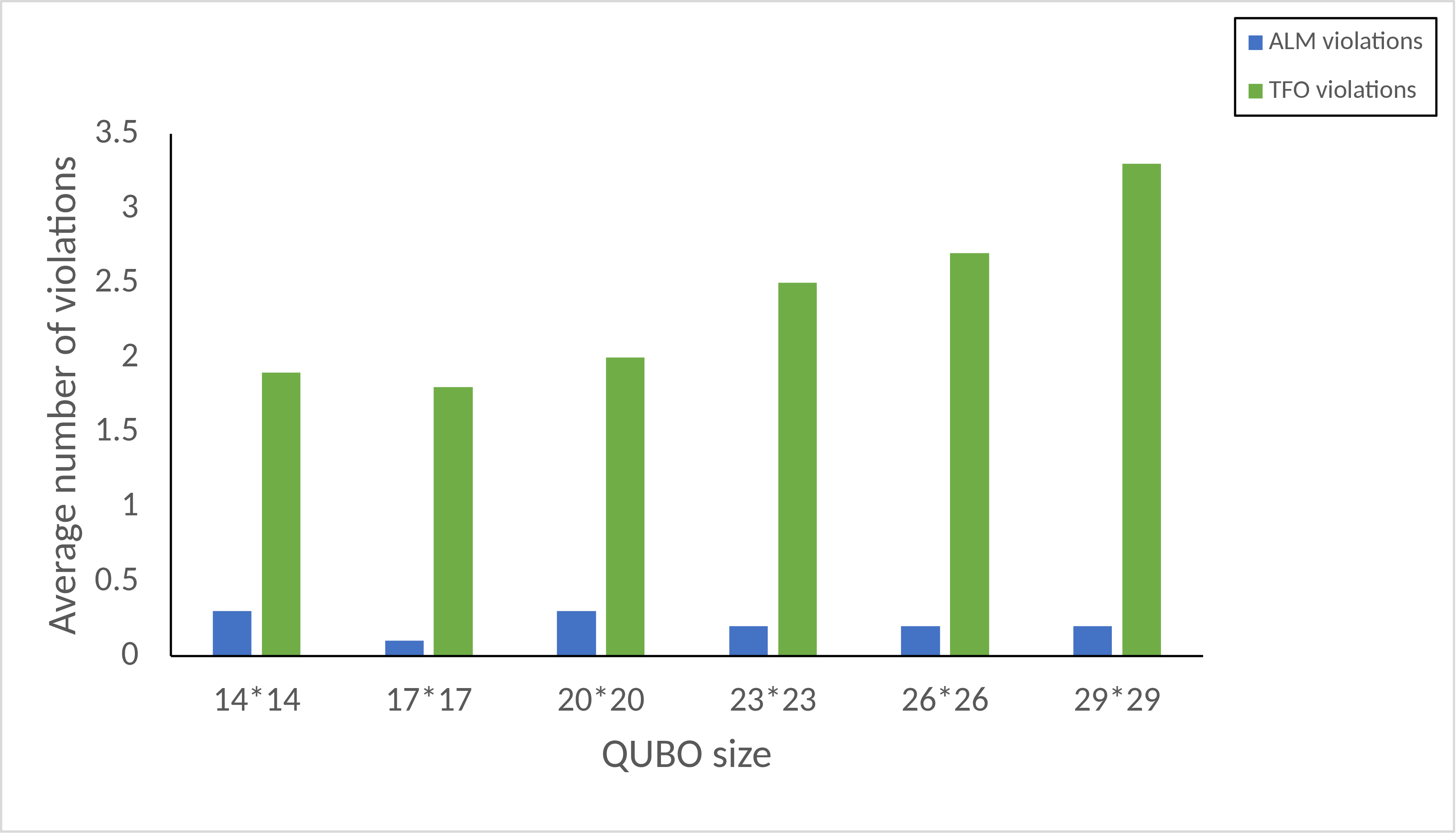}
\caption{Comparison of average number of violations on DWaveSampler(Default Embedding) upon varying QUBO sizes} \label{fig10}
\end{figure}

\begin{figure}
\includegraphics[width=0.5\textwidth]{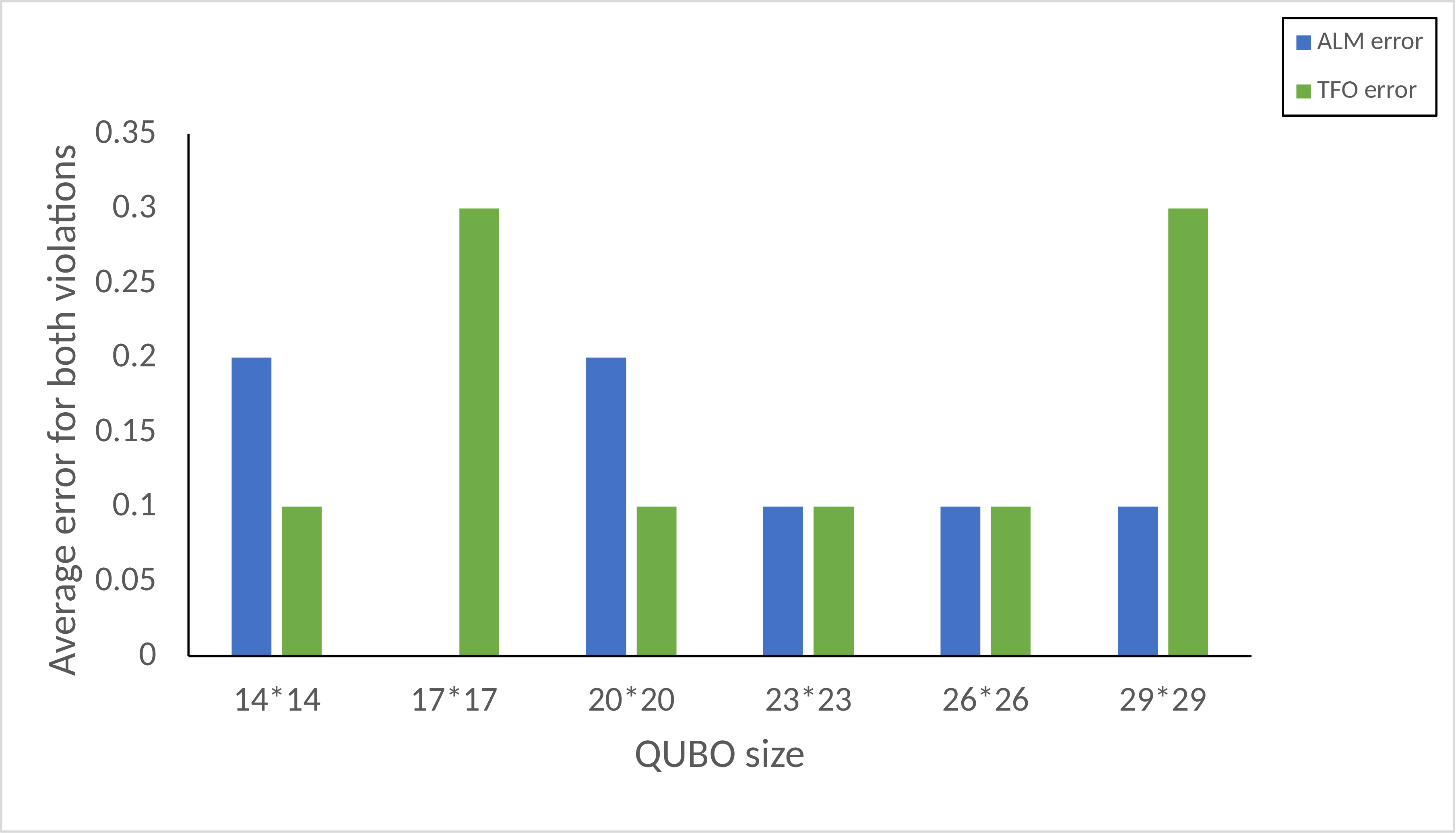}
\caption{Comparison of average error for both types of violations on DWaveSampler(Default Embedding) upon varying QUBO sizes} \label{fig11}
\end{figure}

Here, findings from experiments combining ALM with different number of variables (or changing the size of problem) in TFO are discussed (refer to Figures~\ref{fig7},~\ref{fig8},~\ref{fig9},~\ref{fig10} and~\ref{fig11} for this section). Keeping the number of variables in ALM the same as 5, the number of TFO variables were systematically varied to create combined QUBOs of sizes 14, 17, 20, 23, 26, and 29 variables.

The experimental results revealed a consistent trend across different problem sizes: the parallel solutions obtained through quantum annealing were not optimal when compared to their non-parallel counterparts. Despite varying the problem size, the Parallel Quantum Annealing approach faced challenges in achieving optimal solutions for the combined ALM and TFO problems. Further investigation revealed that even when the ALM and TFO problems were run independently on the D-Wave solver with Default Embedding, the results were not consistently the best optimized solutions. This observation suggests that the challenges in achieving optimal solutions might not be solely attributed to the parallel processing aspect, but rather possibly could be due to the size of the problem that can be handled by DWaveSampler itself \cite{willsch2022benchmarking}. On the other hand, the $TTS$ observed for parallel processing was consistently lesser than that of non-parallel processing which suggests faster problem solving.

\subsection{DWaveSampler (Custom Embedding):}

Keeping the solver same, the embedding technique was changed to custom embedding using hardware information about the D-Wave Advantage System 4.1 \cite{PhysRevA.92.042310, okada2019improving}. Considering the combination of the previously mentioned problems of ALM and TFO:

\subsubsection{Running for 26 variables:}

The experiments revealed that usage of custom embedding improved upon the solution presented by using default embedding. In terms of $SQV$, the solutions were closer to optimal. Across multiple runs, the consistency had also improved as shown by the Standard Deviation of $SQV$. Still the effect of the magnitude disparity between ALM and TFO remained. In terms of $TTS$, the solutions were obtained much faster than their non-parallel counterparts. It also showed a lower degree of average error.

\subsubsection{Effect of problem size:}

\begin{figure}
\includegraphics[width=0.5\textwidth]{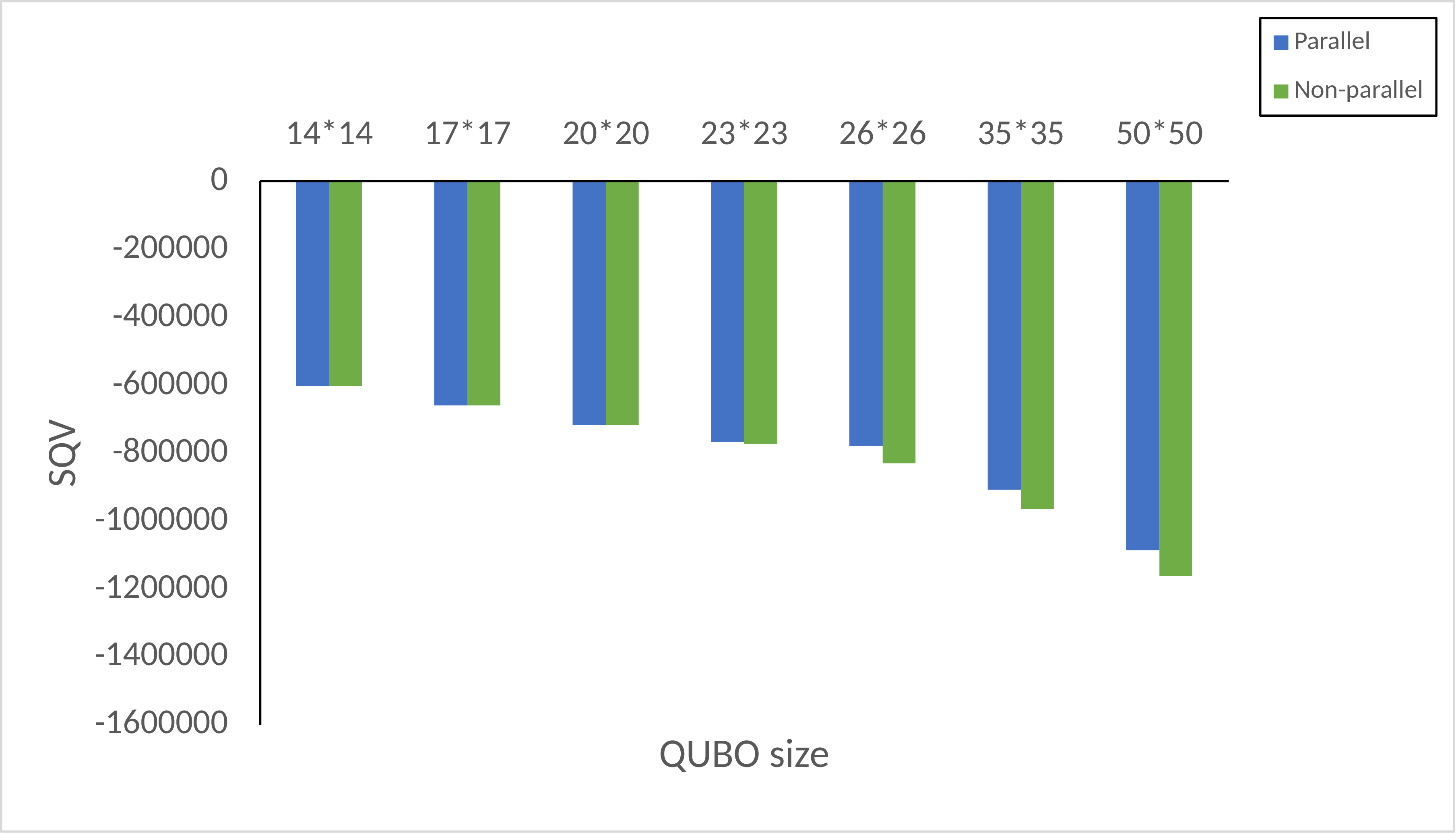}
\caption{Comparison of SQV on DWaveSampler(Custom Embedding) upon varying QUBO sizes} \label{fig12}
\end{figure}

\begin{figure}
\includegraphics[width=0.5\textwidth]{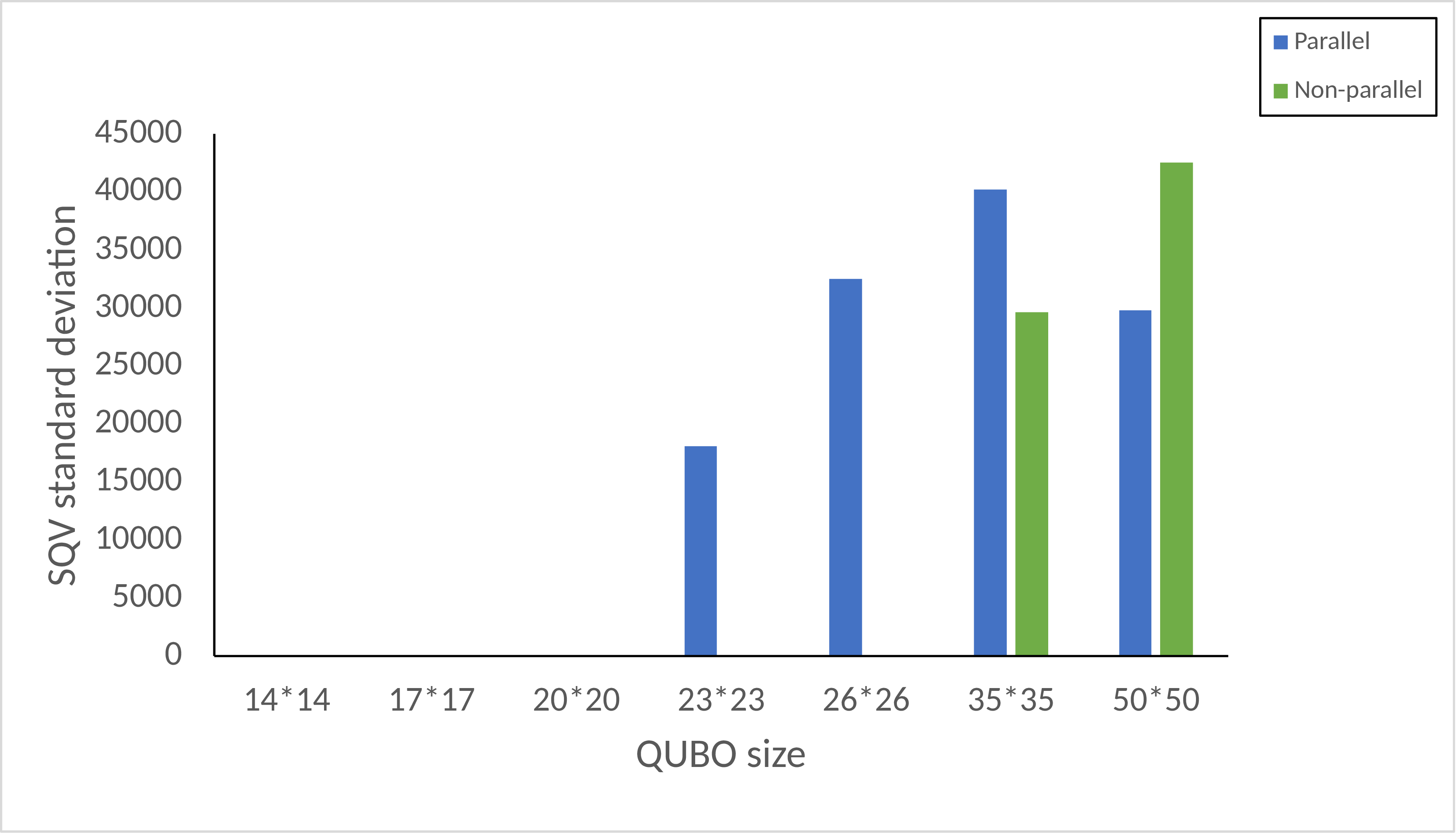}
\caption{Comparison of variation in solution on DWaveSampler(Custom Embedding) upon varying QUBO sizes} \label{fig13}
\end{figure}

\begin{figure}
\includegraphics[width=0.5\textwidth]{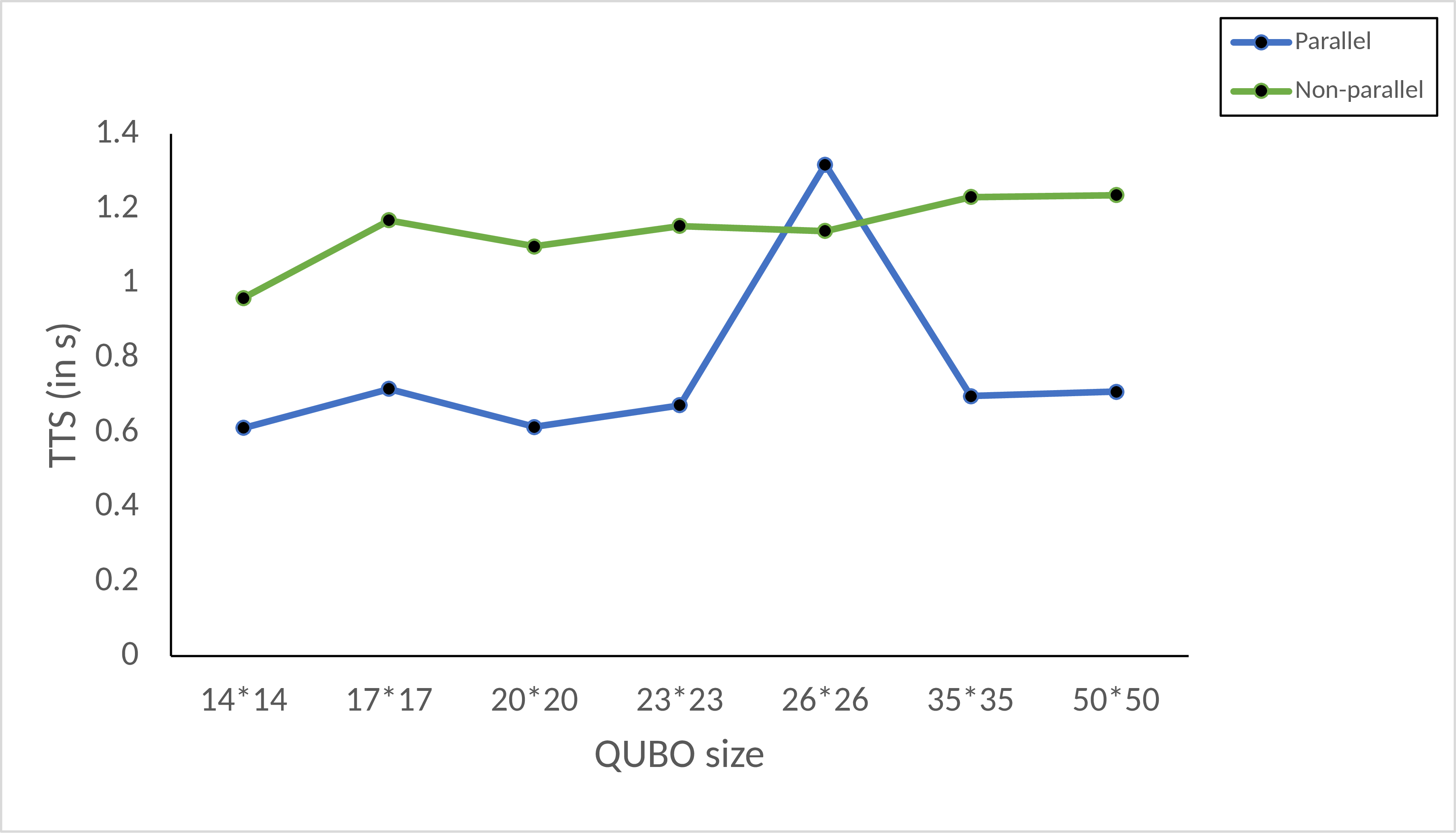}
\caption{Comparison of TTS on DWaveSampler(Custom Embedding) upon varying QUBO sizes} \label{fig14}
\end{figure}

\begin{figure}
\includegraphics[width=0.5\textwidth]{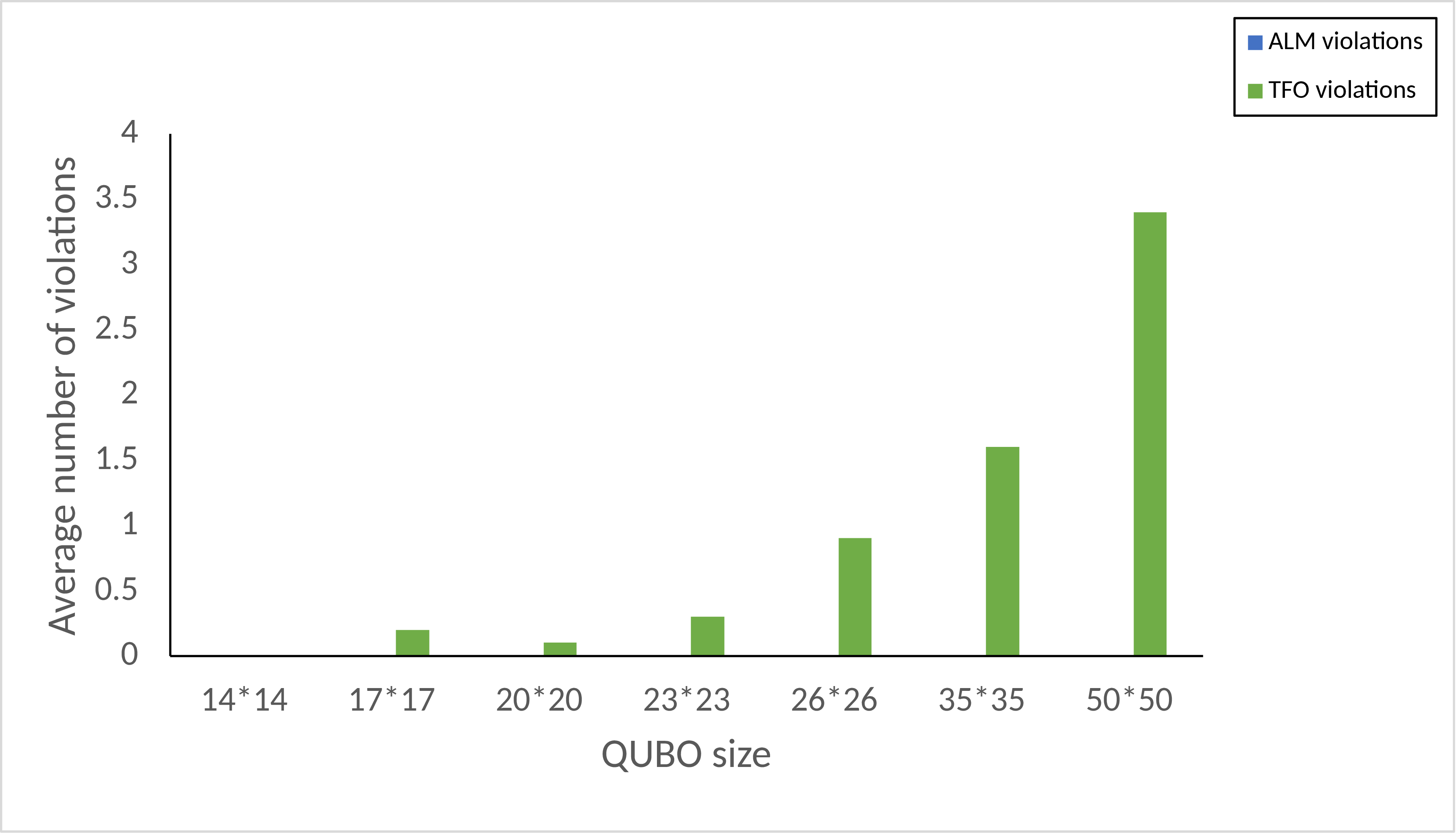}
\caption{Comparison of average number of violations on DWaveSampler(Custom Embedding) upon varying QUBO sizes} \label{fig15}
\end{figure}

\begin{figure}
\includegraphics[width=0.5\textwidth]{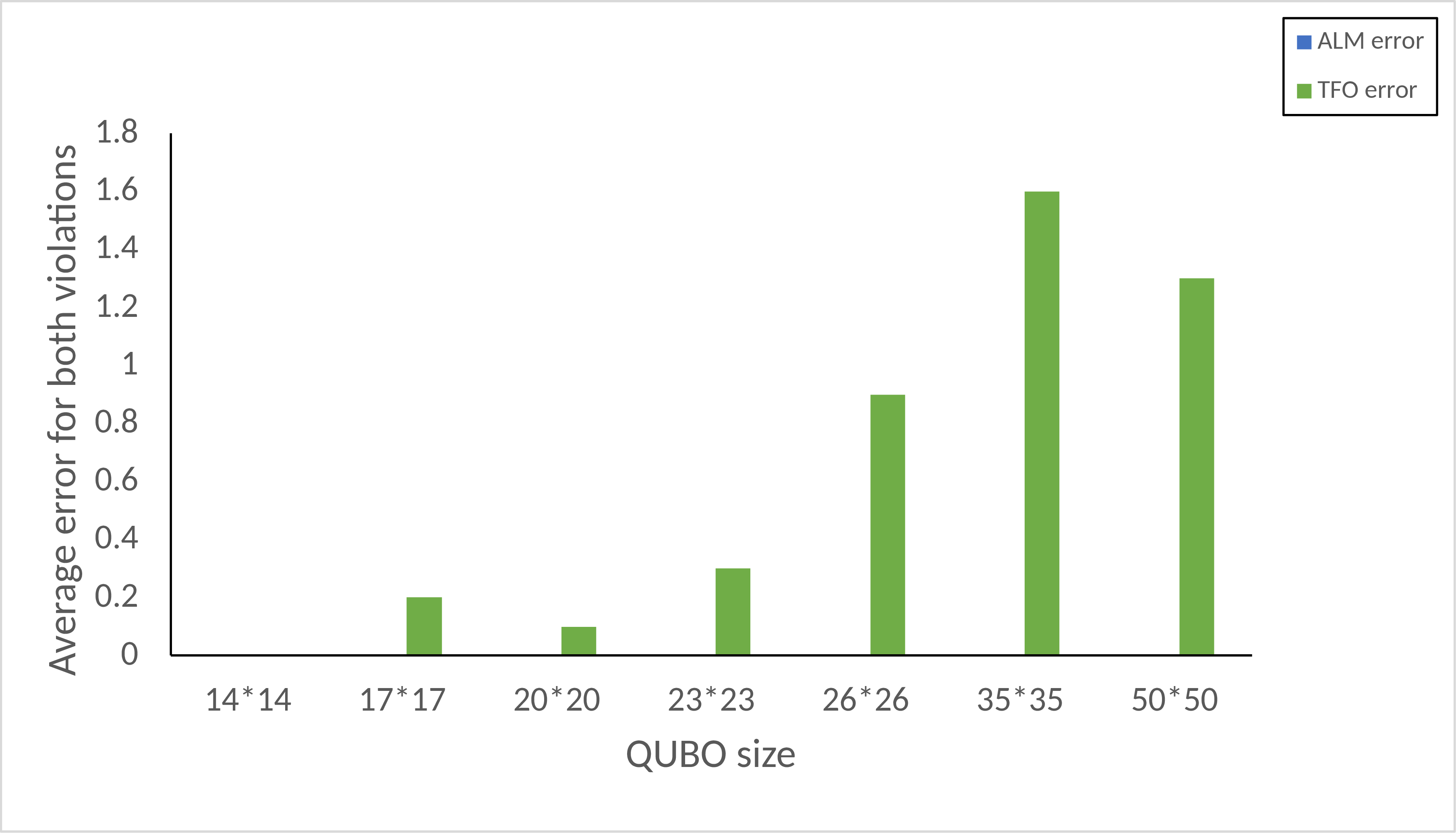}
\caption{Comparison of average error for both types of violations on DWaveSampler(Custom Embedding) upon varying QUBO sizes} \label{fig16}
\end{figure}

Considering the constraints on the size of problems that can be run on a pure Quantum Annealer, the problem size was decreased to check the effectiveness of the Custom Embedding method to quickly find the ground state (refer to Figures~\ref{fig12},~\ref{fig13},~\ref{fig14},~\ref{fig15} and~\ref{fig16} for this section)

Upon reducing the number of variables, and thereby reducing the QUBO size, it was observed that the results were improved. Specifically, for QUBO size of 14, we find that the results are ideal. $SQV$ is optimal and there are no violations with an improvement of $TTS$ over the non-classical runs. As we further increase QUBO size, the results become near-optimal staying very close to their optimal until a threshold of QUBO size = 23. Until this threshold, all metrics are very close to their optimal values with the Standard Deviation of $SQV$ being $\approx$ 0. Beyond this point, we observe that even though we get a benefit of run time, the solution results are not that close to their non-parallel counterparts in terms of all the documented metrics. The difference between the parallel and non-parallel solutions in terms of both $SQV$ and Average error of violations from QUBO size = 26 only started to increase as the QUBO size increased and the parallel solutions became increasingly more inconsistent as shown by the increasing Standard Deviation of $SQV$. The run time benefit over the non-parallel solution still remained. 

\subsubsection{Using Normalization for original problem:}

\begin{figure}
\includegraphics[width=0.5\textwidth]{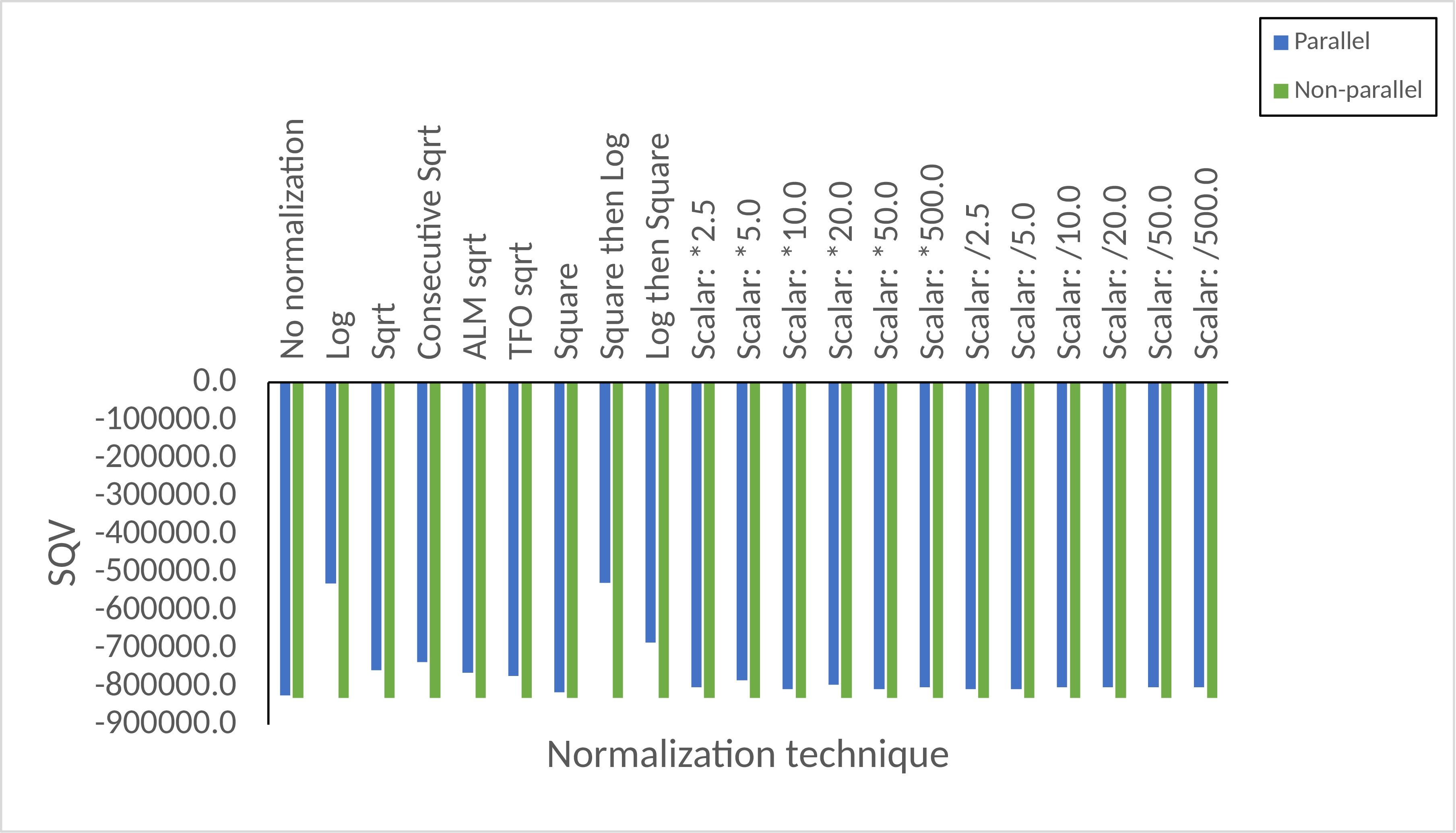}
\caption{Comparison of SQV for normalization techniques on DWaveSampler(Custom Embedding) for QUBO Size: 26×26} \label{fig17}
\end{figure}

\begin{figure}
\includegraphics[width=0.5\textwidth]{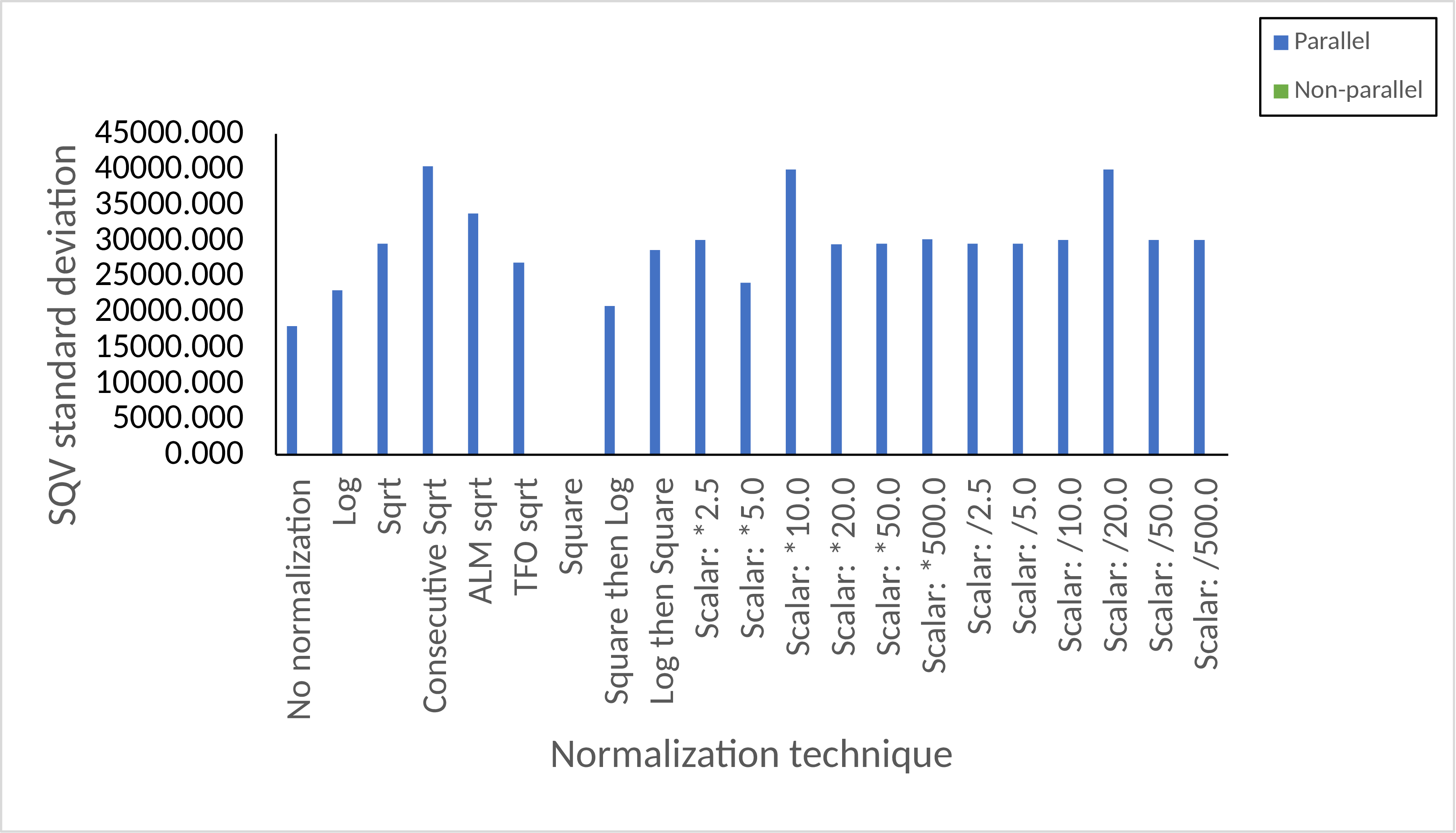}
\caption{Comparison of variation in solution for normalization techniques on DWaveSampler(Custom Embedding) for QUBO Size: 26×26} \label{fig18}
\end{figure}

\begin{figure}
\includegraphics[width=0.5\textwidth]{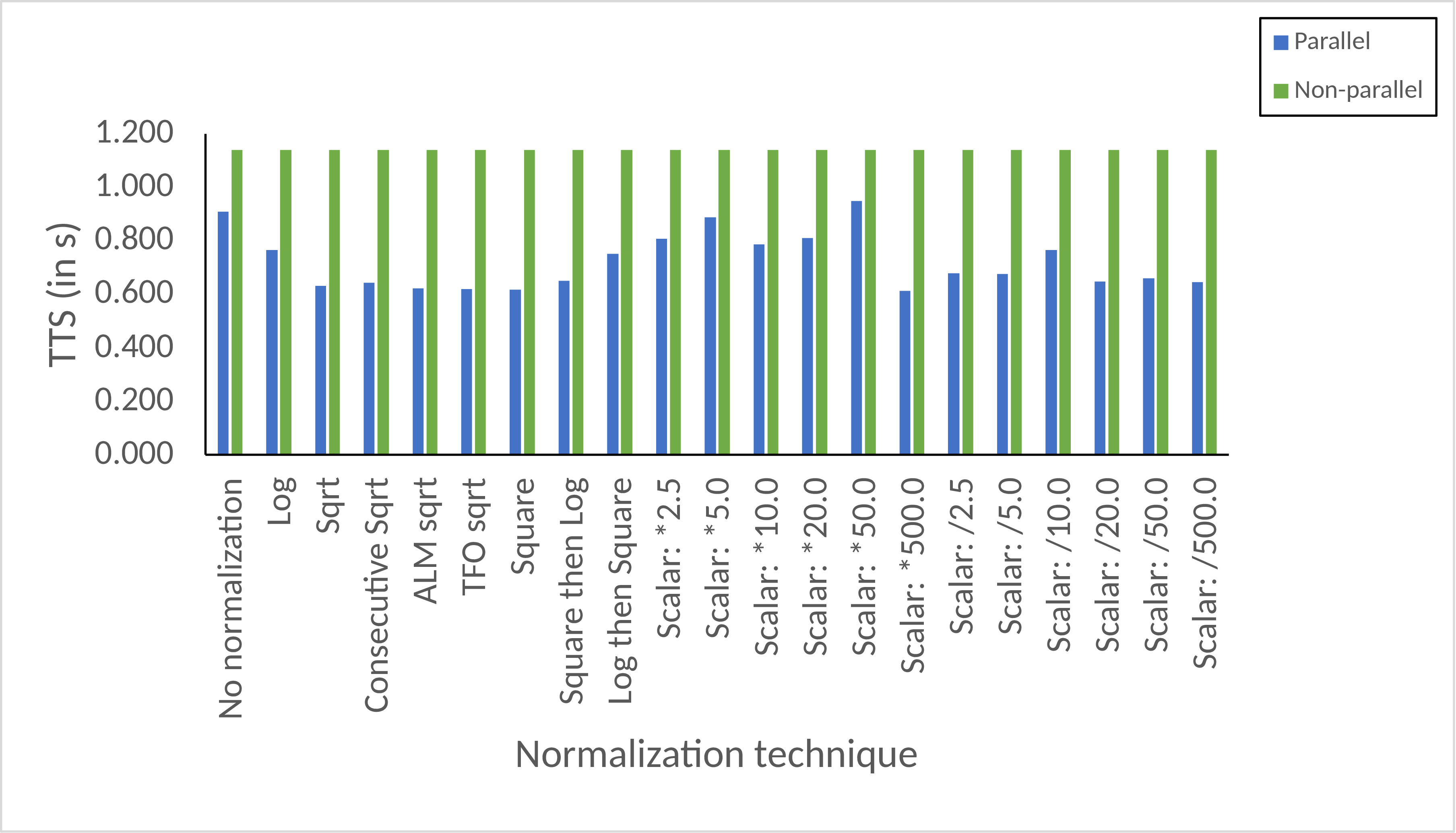}
\caption{Comparison of TTS for normalization techniques on DWaveSampler(Custom Embedding) for QUBO Size: 26×26} \label{fig19}
\end{figure}

\begin{figure}
\includegraphics[width=0.5\textwidth]{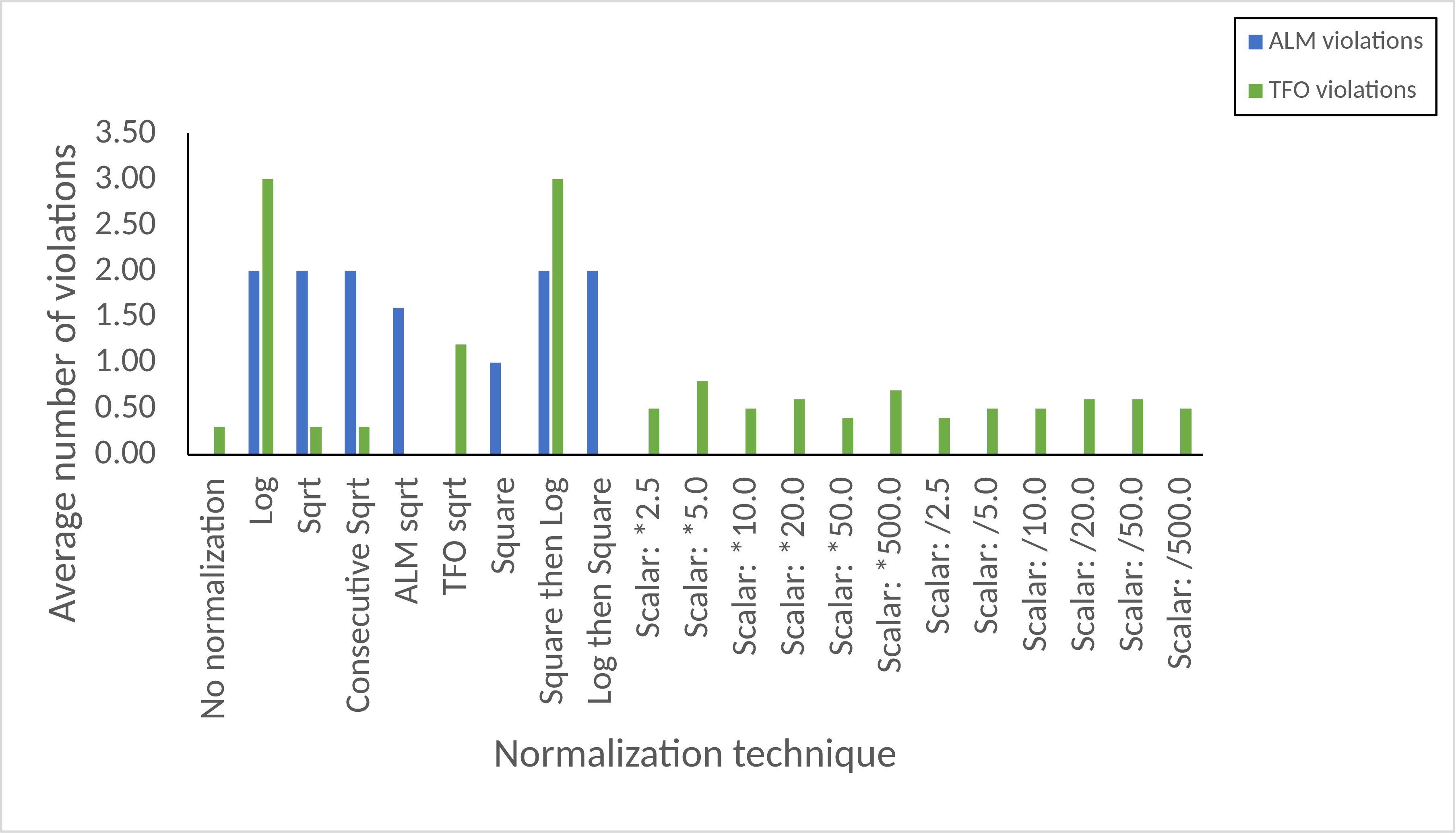}
\caption{Comparison of average number of violations for normalization techniques on DWaveSampler(Custom Embedding) for QUBO Size: 26×26} \label{fig20}
\end{figure}

\begin{figure}
\includegraphics[width=0.5\textwidth]{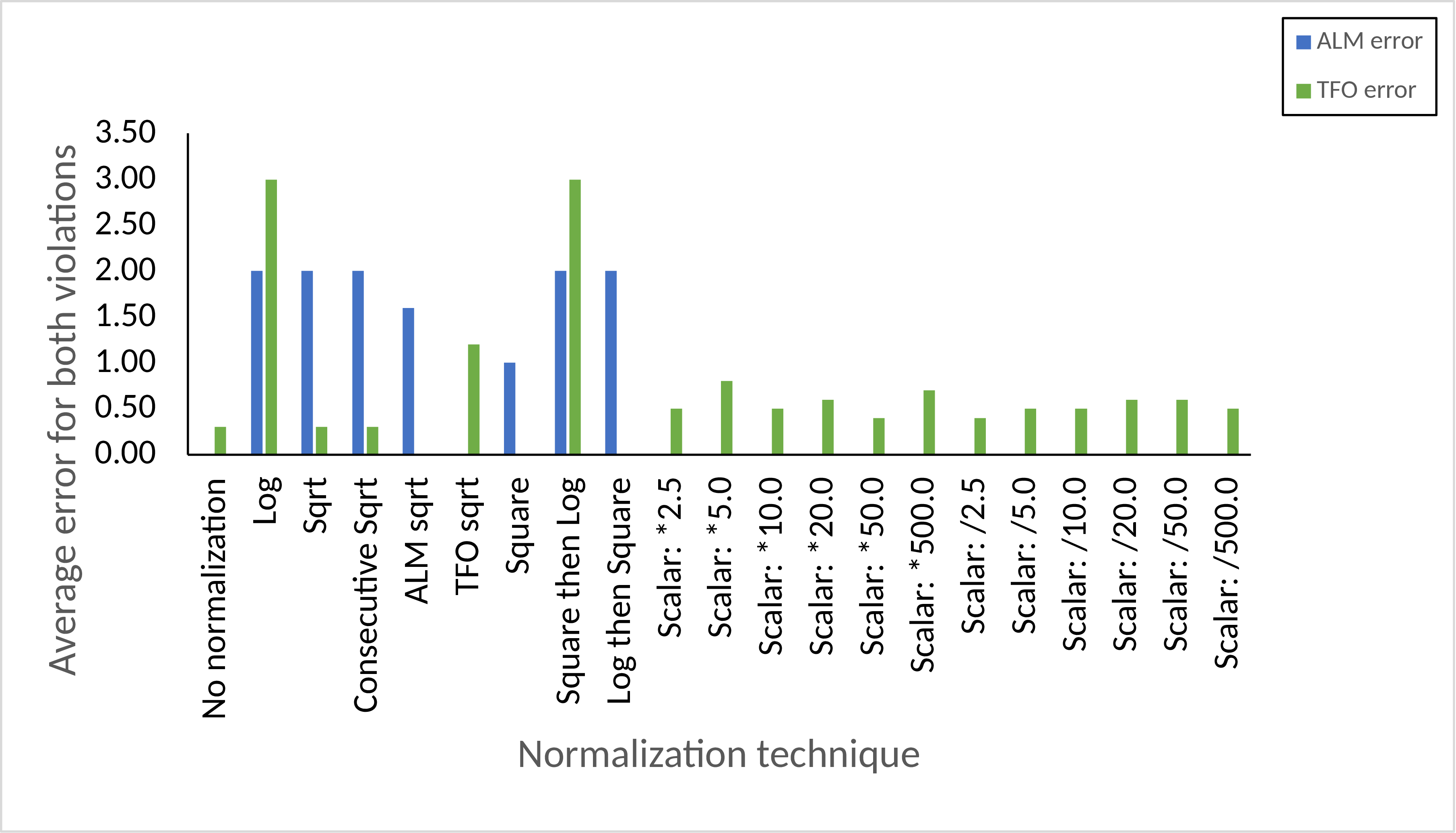}
\caption{Comparison of average error for both types of violations for normalization techniques on DWaveSampler(Custom Embedding) for QUBO Size: 26×26} \label{fig21}
\end{figure}

For improving the solution quality for the experiment with number of variables = 26, different normalization techniques were experimented with to gauge whether they could improve the results of custom embedding (refer to Figures~\ref{fig17},~\ref{fig18},~\ref{fig19},~\ref{fig20} and~\ref{fig21} for this section). 

Again, contrary to expectations, none of the methods used seemed to improve the results considerably as compared to their counterpart with no normalization. Out of all the methods, the scalar normalization seemed to work the best with it providing similar $SQV$ and Total Average number of violations of both the errors combined. Logarithmic normalization seemed to perform the worst since it heavily reduced the magnitude difference, much more than that required to possibly distinguish the two problems to be solved correctly independently. For scalar normalization, the standard deviation also remained mostly consistent.

\subsection{LeapHybridSampler (Default Embedding):}

For obtaining better optimized solution quality and computational efficiency within the Parallel Quantum Annealing framework, the implementation of the LeapHybridSampler was explored \cite{boothby2020nextgeneration, math10081294}.

\begin{figure}
\includegraphics[width=0.5\textwidth]{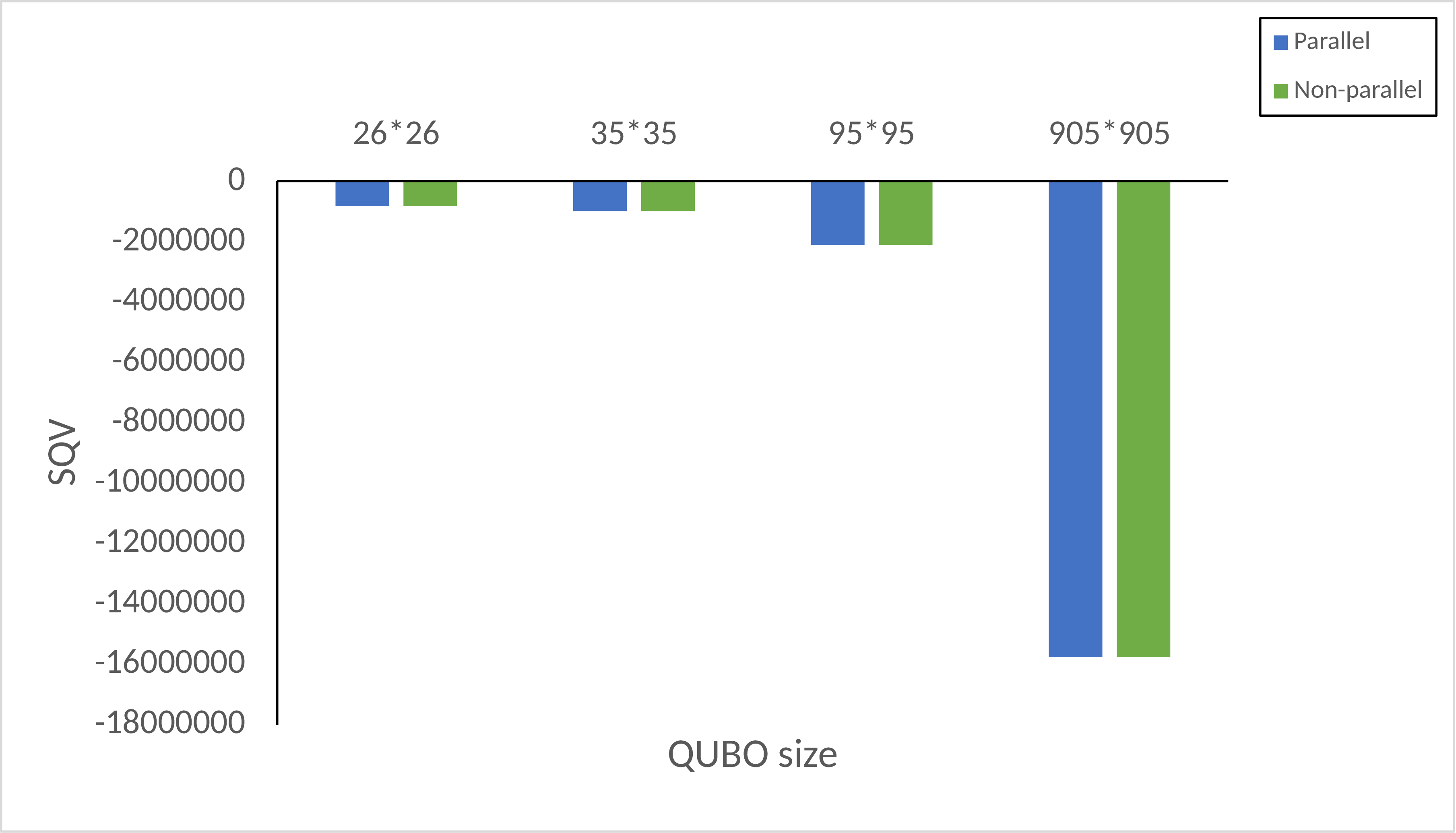}
\caption{Comparison of SQV on LeapHybridSampler(Default Embedding) upon varying QUBO sizes} \label{fig22}
\end{figure}

\begin{figure}
\includegraphics[width=0.5\textwidth]{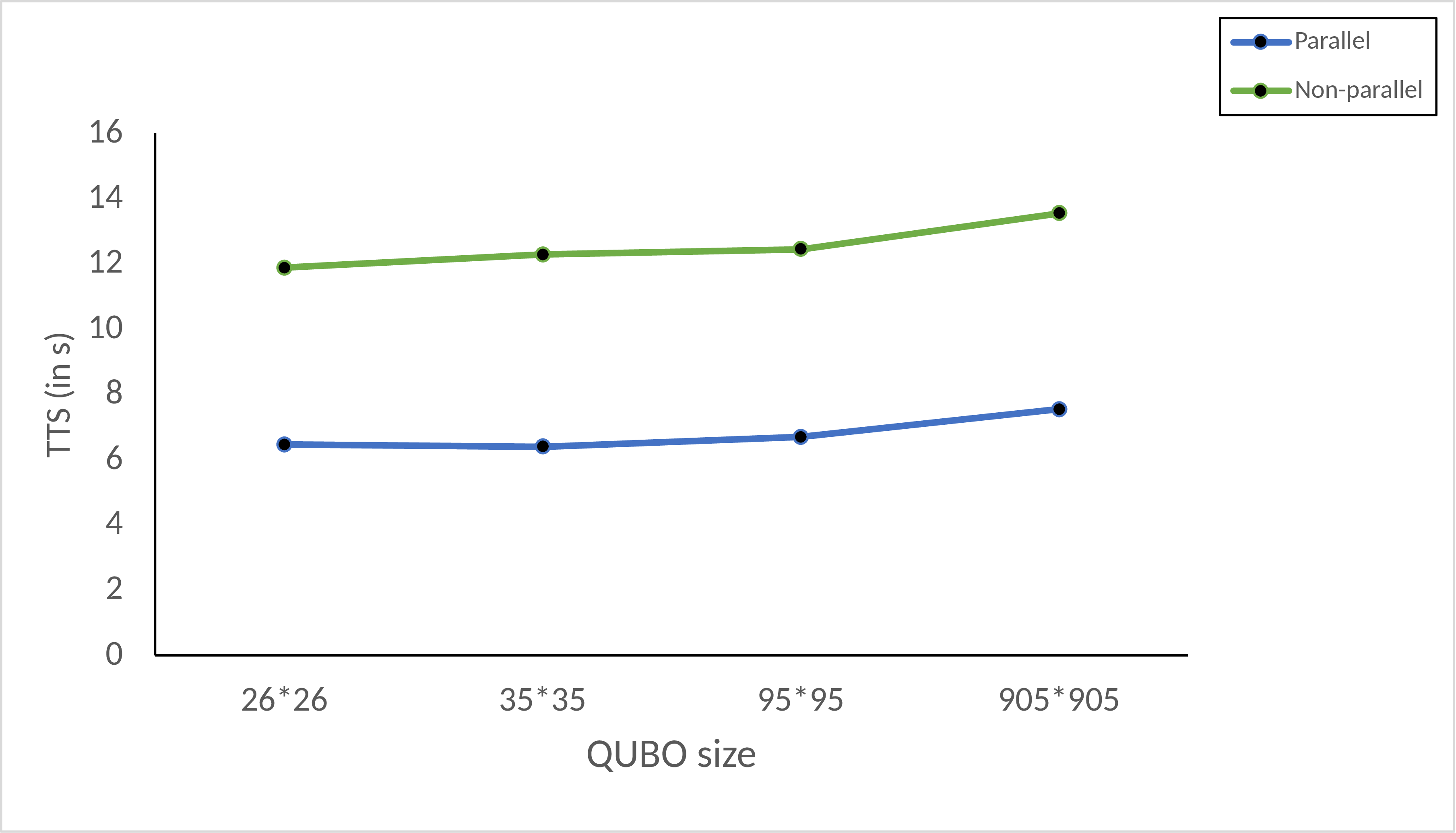}
\caption{Comparison of TTS on LeapHybridSampler(Default Embedding) upon varying QUBO sizes} \label{fig23}
\end{figure}

\begin{figure}
\includegraphics[width=0.5\textwidth]{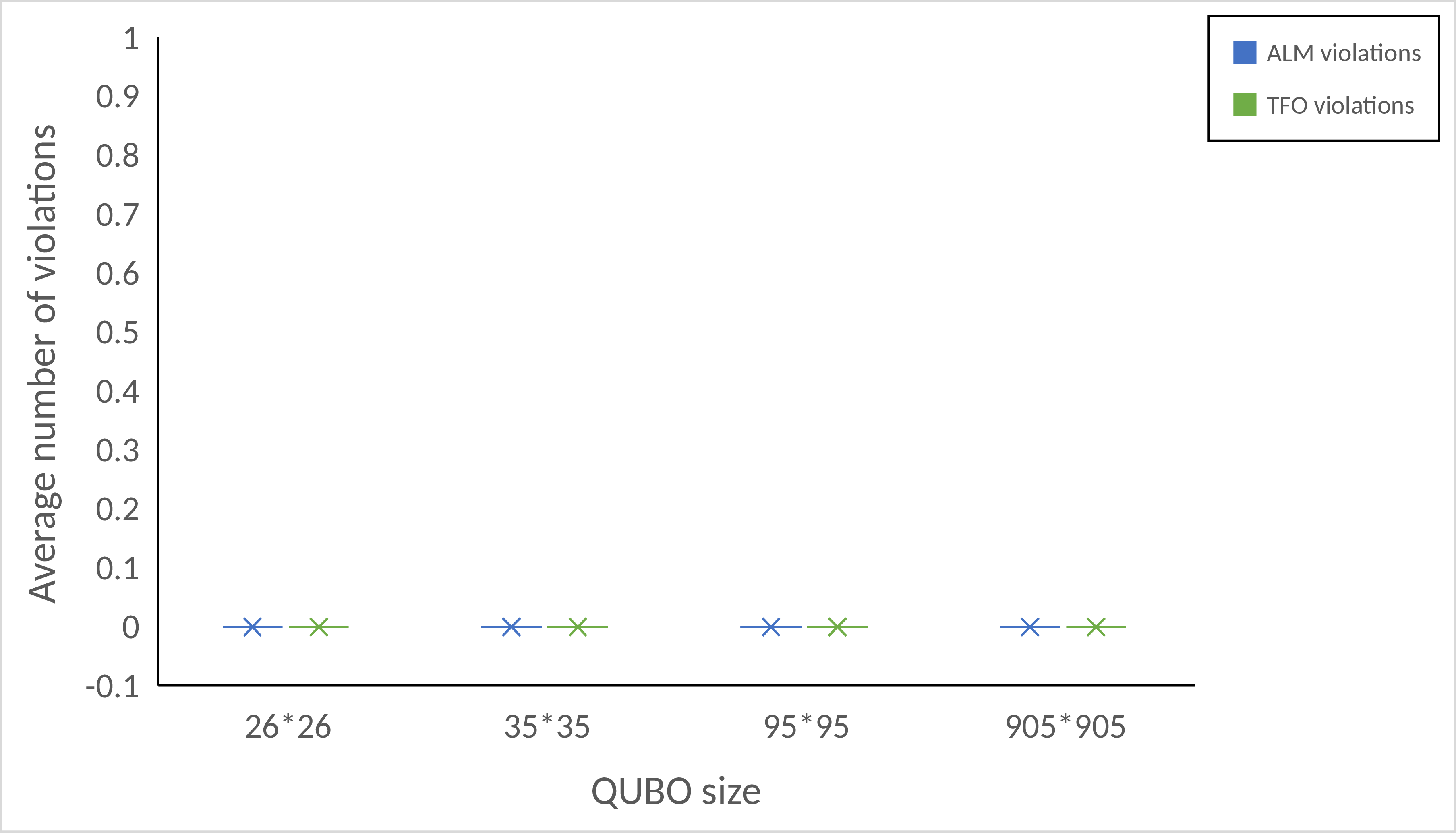}
\caption{Comparison of average number of violations on LeapHybridSampler(Default Embedding) upon varying QUBO sizes} \label{fig24}
\end{figure}

To assess the effectiveness of the LeapHybridSampler, the ALM problem's five variables (with coefficients of magnitude $10^5$) with varying number of variables in the TFO problem (with coefficients of magnitude $10^4$) were combined. The number of TFO variables were varied systematically to create combined QUBOs of sizes 26, 35, 95, and 905 variables (refer to Figures~\ref{fig22},~\ref{fig23} and~\ref{fig24} for this section).

Remarkably, across all problem sizes tested, the LeapHybridSampler consistently yielded the best-case results in the parallel scenario when compared to the non-parallel counterpart. This outcome marked a significant advancement, as the LeapHybridSampler demonstrated its capacity to outperform the other solvers while leveraging the benefits of parallel quantum annealing.

An intriguing observation was the reduction in time to solution achieved through parallelization with the LeapHybridSampler. This efficiency gain stemmed from the characteristics of the D-Wave hybrid solver's sampling time, which behaves as a step function for certain variable sizes. Consequently, the overall time taken for annealing was considerably reduced, enhancing computational efficiency. This was tested to around a higher number of variables $\approx$ 900 as shown.

Notably, the experiments revealed that the LeapHybridSampler consistently produced results without violations of constraints. This underscored the reliability and robustness of the solver in maintaining the integrity of solution outcomes while solving complex optimization problems.

\section{Conclusion}

In this study, investigation of the Parallel Quantum Annealing technique was undertaken, focusing on its capacity to concurrently address multiple optimization problems. Through a series of experiments, the research revealed profound insights into the potentials and constraints associated with this parallelization method. The experimental phase which combined two different problems, also included the implementation of normalization methodologies, exploration of varying problem dimensions, and the utilization of the DWaveSampler (Default Embedding), DWaveSampler (Custom Embedding) and LeapHybridSampler. 

The combination of two distinct problems, different in magnitude and different in variable sizes, was subjected to parallel execution using the three techniques. The findings from these experiments indicate that the default embedding in the DWaveSampler yielded suboptimal results, particularly for the problem with lower magnitude. Even after normalization, the default embedding's performance remained suboptimal. Experimenting with different variable sizes showed improved results for smaller problem sizes, but the improvement was still not optimal.

The custom embedding in the DWaveSampler outperformed the default embedding but it also wasn't suitable for a variable count exceeding $\approx$ 26. The effect of altering the number of variables was evident, with optimal outcomes achieved for small problems, $\approx$ 14 variables with mostly reduced run time as compared to their non-parallel coounterparts. However, a threshold was observed beyond which results became considerably non-ideal. Even normalization applied to the QUBO for 26 variables did not yield enhancement in results.

Remarkably, the LeapHybridSampler using default embedding provided ideal solutions similar to the non-parallel approach. The study extended testing to $\approx$ 900 variables, demonstrating the efficacy of this technique. It suggests that LeapHybridSampler could be potentially used for parallelizing multiple problems, with a benefit in run time and a reduction in queue time with the quality of results being similar.

\textbf{Disclaimer:} As of the time of this research, PwC does not have any Joint Business Relationship (JBR) with D-Wave.

\bibliography{PQA}

\end{document}